\documentclass[12pt]{JHEP3}
\usepackage{graphicx}
\usepackage{array}
\usepackage{supertabular}
\usepackage{longtable}

\newcommand{\nn}{\nonumber}
\newcommand{\beqn}{\begin{eqnarray}}
\newcommand{\eeqn}{\end{eqnarray}}
\newcommand{\be}{\begin{equation}}
\newcommand{\ee}{\end{equation}}
\newcommand{\eqn}[1]{(\ref{#1})}

\newcommand{\ba}{\begin{array}{c}}
\newcommand{\bat}{\begin{array}{cc}}
\newcommand{\ea}{\end{array}}

\newcommand{\bi}{\begin{itemize}}
\newcommand{\ei}{\end{itemize}}
\newcommand{\chpt}{$\chi$PT}
\newcommand{\rcht}{R$\chi$T}

\newcommand{\Frac}[2]{\frac{\displaystyle #1}{\displaystyle #2}}
\newcommand{\no}{\nonumber}

\newcommand{\imag}{{\rm Im}}
\newcommand{\cO}{{\cal O}}

\newcommand{\mbf}{\mathbf}

\newcommand{\bea}{\begin{eqnarray}}
\newcommand{\eea}{\end{eqnarray}}
\newcommand{\beq}{\begin{equation}}
\newcommand{\eeq}{\end{equation}}

\newcommand{\Int}{\displaystyle{\int}}

\newcommand{\bear}{\begin{eqnarray}}
\newcommand{\eear}{\end{eqnarray}}
\newcommand{\mF}{\mathcal{F}}
\newcommand{\mG}{\mathcal{G}}

\newcommand{\ket}{\,\rangle}
\newcommand{\bra}{\langle \,}

\title{Form-factors and current correlators: chiral couplings
$\mathbf{L_{10}^r(\mu)}$ and $\mathbf{C_{87}^r(\mu)}$  at NLO in $1/N_C$
}

\author{Antonio Pich,$^a$ Ignasi Rosell$^{ab}$ and Juan Jos\'e Sanz-Cillero$^c$ \\
$^a$Departament de F\'\i sica Te\`orica, IFIC, Universitat de Val\`encia - CSIC \\
Apt. Correus 22085, E-46071 Val\`encia, Spain \\
$^b$Departamento de Ciencias F\'\i sicas, Matem\'aticas y de la Computaci\'on, \\
Universidad CEU Cardenal Herrera, c/ Sant Bartomeu 55, \\
E-46115 Alfara del Patriarca, Val\`encia, Spain \\
$^c$Grup de F\'\i sica Te\`orica and IFAE \\
Universitat Aut\`onoma de Barcelona, E-08193 Barcelona, Spain \\
E-mail: \email{Antonio.Pich@uv.es,rosell@uch.ceu.es,cillero@ifae.es}}

\abstract{
Using the resonance chiral theory Lagrangian, we perform a calculation of the vector and axial-vector two-point functions at the next-to-leading order (NLO) in the $1/N_C$ expansion. We have analyzed these correlators within the
single-resonance approximation and have also investigated the corrections induced by a second multiplet of vector and axial-vector resonance states. Imposing the correct QCD short-distance constraints, one determines the difference of the two correlators $\Pi(t)\equiv\Pi_{VV}(t)-\Pi_{AA}(t)$ in terms of the pion decay constant and resonance masses. Its low momentum expansion fixes then the low-energy chiral couplings $L_{10}$ and $C_{87}$ at NLO, keeping full control of their renormalization scale dependence. At $\mu_0=0.77$~GeV, we obtain $L_{10}^r(\mu_0) = (-4.4 \pm 0.9)\cdot 10^{-3}$ and $C_{87}^r(\mu_0)=(3.1 \pm 1.1)\cdot 10^{-5}$.}


\preprint{IFIC/08$-$14\\ FTUV/08$-$0311 \\}

\begin{document}





\section{Introduction}

Effective field theory \cite{EFT} is nowadays the standard technique to investigate the low-energy dynamics of QCD. In particular, the chiral symmetry constraints encoded in Chiral Perturbation Theory (\chpt) provide a very powerful tool
to access the non-perturbative regime through a perturbative expansion in powers of light quark masses and momenta \cite{Weinberg,ChPTp4,ChPTrevs}. The precision required in present phenomenological applications makes necessary to include corrections of $\cO(p^6)$. While many two-loop \chpt\ calculations have been already carried out~\cite{ChPTp6,review},
the large number of unknown low-energy couplings (LECs) appearing at this order puts a clear limit to the achievable accuracy \cite{review}. A dynamical determination of these \chpt\ couplings is compulsory to achieve further progress in our understanding of strong interactions at low energies.

A useful connection between \chpt\ and the underlying QCD dynamics can be established in the limit of an infinite number of quark colours \cite{polychromatic,MHA}. Assuming confinement, the strong dynamics at $N_C\to\infty$ is given by tree diagrams with infinite sums of hadron exchanges, which correspond to the tree approximation of some local effective Lagrangian \cite{NC}. Resonance Chiral Theory (\rcht) provides a correct framework to incorporate these massive mesonic states within an effective Lagrangian formalism \cite{RChTa,RChTb,RChTc}. Integrating out the heavy fields one recovers at low energies the \chpt\ Lagrangian with explicit values of the chiral LECs in terms of resonance parameters. Since the short-distance properties of QCD impose stringent constraints on the \rcht\ couplings, it is then possible to extract information on the low-energy \chpt\ parameters.

Clearly, we cannot determine at present the infinite number of meson couplings which characterize the large--$N_C$ Lagrangian. This would be equivalent to achieve an explicit dynamical solution of the QCD spectrum in the $N_C\to\infty$ limit. However, one can obtain useful approximations in terms of a finite number of meson fields. Truncating the infinite tower of meson resonances to the lowest states with $0^{-+}$, $0^{++}$, $1^{--}$ and $1^{++}$ quantum numbers (Single-Resonance Approximation), one gets a very successful prediction of the $\cO(p^4 N_C)$ \chpt\ couplings in terms of only three parameters: $M_V$, $M_S$ and the pion decay constant $F$ \cite{polychromatic}. Some $\cO(p^6)$ LECs have been already predicted in this way, by studying an appropriate set of three-point functions \cite{three,VAP}. More recently, the program to determine all $\cO(p^6)$ LECs at leading order in $1/N_C$ has been put on very solid grounds, with a complete classification of the needed terms in the \rcht\ Lagrangian \cite{RChTc}.

Since chiral loop corrections are of next-to-leading order (NLO) in the $1/N_C$ expansion, the large--$N_C$ determination of the LECs is unable to control their renormalization-scale dependence. This introduces unavoidable
theoretical uncertainties, which are very important for couplings related with the scalar sector.
First analyses of resonance loop contributions to the running of $L_{10}^r(\mu)$ and $L_9^r(\mu)$ were attempted in Ref.~\cite{CP:02} and Ref.~\cite{RSP:05} respectively. In spite of all the complexity associated with the still not so well understood renormalization of \rcht\ \cite{RSP:05,RPP:05,natxo-tesis}, these pionnering calculations have shown the potential
predictability at the NLO in $1/N_C$.

Using analyticity and unitarity we can avoid all technicalities associated with the renormalization procedure, reducing the calculation of simple Green functions to tree-level diagrams plus dispersion relations. This allows to understand the underlying physics in a much more transparent way. In particular, the subtle cancellations among many unknown renormalized couplings found in Ref.~\cite{RSP:05} and the relative simplicity of the final result can be better understood in terms of the imposed short-distance constraints. Following these ideas, in Ref.~\cite{L8-nlo} we determined the couplings $L_8^r(\mu)$ and $C_{38}^r(\mu)$ at NLO in $1/N_C$, through an analysis of the difference between the scalar and pseudoscalar current correlators. As a next step, we present in this article the more involved study of the two-point function of a left-handed and a right-handed vector currents, which allows us to perform a NLO determination of the couplings $L_{10}^r(\mu)$ and $C_{87}^r(\mu)$.

To fix the notation, we introduce the \rcht\ Lagrangian in the next section. The current-current correlators are defined in section~\ref{sec:ff}, where we discuss the relation between their absorptive parts and meson form factors. Our study of the vector and axial-vector correlators is presented in section~\ref{sec:v-a}, while section~\ref{sec:L10}
contains the determination of $L_{10}^r(\mu)$ and $C_{87}^r(\mu)$. A summary of our results is finally given in section~\ref{sec:summary}. In order to achieve our goal, we have performed an exhaustive analysis of scalar, pseudoscalar, vector and axial-vector form factors with two external meson legs as final states. The results of this lengthy calculation are given in appendix~\ref{ap:A}. Further technical details on the computation of current-current correlators and their dispersive representation are contained in appendices~\ref{ap:B} and \ref{ap:C}.

\section{The \rcht\ Lagrangian}\label{sec:lagrangian}

Let us adopt the Single Resonance Approximation (SRA), where just the lightest resonances with non-exotic quantum numbers are considered. On account of large-$N_C$, the mesons are put together into $U(3)$ multiplets. Hence, our degrees of freedom are the pseudo-Goldstone bosons (the lightest pseudoscalar mesons)
along with massive multiplets of the type $V(1^{--})$, $A(1^{++})$, $S(0^{++})$ and $P(0^{-+})$. With them, we construct the most general effective action that preserves chiral symmetry. Since we are interested on the structure of the interaction at short distances, we will work in the chiral limit. With this simplification we do not loose any information on the LECs we want to determine, because they are independent of the light quark masses.

The effective Resonance Chiral Theory must satisfy the high-energy behaviour dictated by QCD. To comply with this requirement we will only consider operators constructed with chiral tensors of $\cO(p^2)$; interactions with higher-order chiral tensors would violate the QCD asymptotic behaviour, unless their couplings are severely fine tuned to ensure the needed cancellations at large momenta. Moreover, we will restrict our analysis to operators with a maximum of three resonance fields, because these are the only ones contributing to the observables we are interested in (tree-level two-body form factors and one-loop correlators).

The different terms in the Lagrangian can be classified by their number of resonance fields:
\begin{eqnarray} \label{lagrangian}
\mathcal{L}_{R\chi T}&=&\mathcal{L}_\chi \,+\,\sum_{R_1}\mathcal{L}_{R_1}
\,+\,\sum_{R_1,R_2}\mathcal{L}_{R_1R_2}\,
+\,\sum_{R_1,R_2,R_3}\mathcal{L}_{R_1R_2R_3}\, + \, ... \,\,\,  ,
\end{eqnarray}
where the dots denote the irrelevant operators with four or more resonance fields, and the indices $R_i$ run over all different resonance multiplets, $V$, $A$, $S$ and $P$. The $\cO(p^2)$ $\chi$PT Lagrangian~\cite{ChPTp4},
\begin{eqnarray}
\mathcal{L}_{\chi} &=\, \Frac{F^2}{4} \bra u_\mu u^\mu + \chi_+ \ket \, ,
\end{eqnarray}
contains the terms with no resonance fields. The second term in \eqn{lagrangian} corresponds to the operators with one massive resonance~\cite{RChTa},
\begin{eqnarray}
\mathcal{L}_V &=&\, \frac{F_V}{2\sqrt{2}} \bra V_{\mu\nu} f^{\mu\nu}_+ \ket \,+\, \frac{i\, G_V}{2\sqrt{2}} \bra V_{\mu\nu} [u^\mu, u^\nu] \ket \, ,  \\
\mathcal{L}_A &=&\, \frac{F_A}{2\sqrt{2}} \bra A_{\mu\nu} f^{\mu\nu}_- \ket\, ,\\
%
\mathcal{L}_S &=&\, c_d \bra S u_\mu u^\mu\ket\,+\,c_m\bra S\chi_+\ket\,  \, , \phantom{\frac{1}{2}}  \\
\mathcal{L}_P &=&\, i\,d_m \bra P \chi_- \ket\, .\phantom{\frac{1}{2}} \label{P}
\end{eqnarray}
The Lagrangian $\mathcal{L}_{R_1R_2}$ contains the kinetic resonance terms and the remaining operators with
two resonance fields~\cite{RChTa,RChTc,RSP:05},
\begin{eqnarray}
\mathcal{L}^{\,\mathrm{kin}}_{RR} &=&\, \frac{1}{2} \bra \nabla^\mu R\nabla_\mu R\,-\, M_R^2 R^2 \ket  \, , \qquad \qquad \, \, \, \, \qquad (R=S,P)  \\
\mathcal{L}^{\,\mathrm{kin}}_{RR} &=&\, -\frac{1}{2} \bra \nabla^\lambda R_{\lambda\mu} \nabla_\nu R^{\nu\mu}\,-\,
\frac{1}{2}M_R^2 R_{\mu\nu}R^{\mu\nu} \ket \, , \quad (R=V,A)    \\
%
\mathcal{L}_{RR}&=&\, \lambda^{RR}_1 \bra RR \,u^\mu u_\mu \ket + \lambda^{RR}_2 \bra R u_\mu R u^\mu  \ket  + \lambda^{RR}_3 \bra RR\, \chi_+ \ket \,,\quad\quad (R=S,P)\, \qquad\phantom{\frac{1}{2}}  \\
\mathcal{L}_{SP}&=&\,\lambda^{SP}_1 \bra u_\alpha \{\nabla^\alpha S, P \} \ket +i \lambda^{SP}_2 \bra \{S,P\} \chi_-\ket \, , \phantom{\frac{1}{2}} \\
%
\mathcal{L}_{SV}&=&\, i \lambda^{SV}_1 \bra \{ S, V_{\mu \nu}\} u^\mu u^\nu \ket \,+\, i\lambda^{SV}_2 \bra S u_\mu V^{\mu\nu} u_\nu \ket \,+\, \lambda^{SV}_3 \bra \{ S, V_{\mu\nu}\} f^{\mu\nu}_+ \ket \, , \phantom{\frac{1}{2}} \\
\mathcal{L}_{SA}&=&\, \lambda^{SA}_1 \bra \{\nabla_\mu S, A^{\mu\nu} \} u_\nu \ket \,+\,  \lambda^{SA}_2 \bra \{ S, A_{\mu\nu} \} f^{\mu\nu}_- \ket \, ,\phantom{\frac{1}{2}}  \\
%
%
\mathcal{L}_{PV}&=&\, i \lambda^{PV}_1\bra [\nabla^\mu P,V_{\mu\nu} ] u^\nu \ket \,+\, i  \lambda^{PV}_2 \bra [P, V_{\mu\nu} ] f^{\mu\nu}_- \ket \, , \phantom{\frac{1}{2}} \\
\mathcal{L}_{PA}&=&\, i \lambda^{PA}_1  \bra [P, A_{\mu\nu} ]f^{\mu\nu}_+ \ket \,+\,\lambda^{PA}_2 \bra [ P, A_{\mu\nu} ] u^\mu u^\nu \ket \, ,\phantom{\frac{1}{2}} \\
\mathcal{L}_{VA}&=&\,  \lambda^{VA}_1\bra [ V^{\mu\nu}, A_{\mu\nu} ] \chi_- \ket \,+\, i \lambda^{VA}_2 \bra  [ V^{\mu\nu}, A_{\nu\alpha} ] h^\alpha_\mu \ket \,+\, i \lambda^{VA}_3 \bra  [ \nabla^\mu V_{\mu\nu}, A^{\nu\alpha} ] u_\alpha \ket\phantom{\frac{1}{2}} \nonumber \\
&&+ i \lambda^{VA}_4 \bra  [ \nabla_\alpha V_{\mu\nu}, A^{\alpha\nu} ] u^\mu \ket \,+\, i \lambda^{VA}_5 \bra  [ \nabla_\alpha V_{\mu\nu}, A^{\mu\nu} ] u^\alpha \ket \phantom{\frac{1}{2}} \nonumber \\ &&+\, i \lambda^{VA}_6 \bra  [  V_{\mu\nu}, A^{\mu}_{\,\,\alpha} ]f^{\alpha\nu}_- \ket  ,\phantom{\frac{1}{2}} \!\!\!\!\!\!\!\!\\
\mathcal{L}_{RR}&=&\,\lambda^{RR}_1\bra R_{\mu\nu} R^{\mu\nu} u_\alpha u^\alpha\ket \,+\,\lambda^{RR}_2 \bra R_{\mu\nu} u^\alpha R^{\mu\nu} u_\alpha \ket  \,+\,\lambda^{RR}_3 \bra R_{\mu\alpha}  R^{\nu\alpha} u^\mu u_\nu \ket  \phantom{\frac{1}{2}}   \nonumber \\
&&+\,\lambda^{RR}_4 \bra R_{\mu\alpha}  R^{\nu\alpha}  u_\nu u^\mu \ket \, +\,\lambda^{RR}_5 \bra R_{\mu\alpha} \left(  u^\alpha R^{\mu\beta}  u_\beta +u_\beta R^{\mu\beta}u^\alpha\right) \ket  \phantom{\frac{1}{2}}\nonumber \\
&& +\,\lambda^{RR}_6 \bra R_{\mu\nu} R^{\mu\nu}  \chi_+ \ket \,+\,i\lambda^{RR}_7 \bra R_{\mu\alpha} R^{\alpha}_{\,\,\nu}f^{\mu\nu}_+ \ket \, . \qquad \qquad (R=V,A)\phantom{\frac{1}{2}}
\end{eqnarray}
Finally, the last piece of Eq.~(\ref{lagrangian}) includes the operators with three resonance fields. We only list those operators needed for the calculation of the form factors we are interested in (terms with only resonance fields and covariant derivatives $\nabla_\mu$):
\begin{eqnarray}
\Delta \mathcal{L}_{SRR}&=&\,\lambda^{SRR}_0 \bra SRR \ket \,+\, \lambda^{SRR}_1 \bra S \,\nabla_\mu R \, \nabla^\mu R \ket \, , \quad\quad (R=S,P) \phantom{\frac{1}{2}} \\
%
%
\Delta \mathcal{L}_{SRR}&=&\lambda^{SRR}_0 \bra S R_{\mu\nu} R^{\mu\nu} \ket + \lambda^{SRR}_1 \bra S \, \nabla_\mu R^{\mu\alpha} \, \nabla^\nu R_{\nu\alpha} \ket \nonumber + \lambda^{SRR}_2 \bra S \, \nabla^\nu R^{\mu\alpha} \, \nabla_\mu R_{\nu\alpha} \ket\phantom{\frac{1}{2}} \\
&&   +\, \lambda^{SRR}_3 \bra S \, \nabla_\alpha R^{\mu\nu} \, \nabla^\alpha R_{\mu\nu} \ket \, +\,\lambda^{SRR}_4 \bra S \{ R^{\mu\nu} , \nabla^2 R_{\mu\nu} \ket \nonumber \\
&& +\,\lambda^{SRR}_5 \bra S \{ R_{\mu\alpha} , \nabla^\mu \nabla_\nu R^{\nu\alpha}\} \ket \, , \qquad \qquad \qquad  (R=V,A)\phantom{\frac{1}{2}} \\
\Delta \mathcal{L}_{SPA}&=&\,\lambda^{SPA} \bra A^{\mu\nu} \{ \nabla_\mu S, \nabla_\nu P \} \ket \, ,  \phantom{\frac{1}{2}}\\
%
\Delta \mathcal{L}_{PVA}&=&\,i \lambda^{PVA}_0 \bra P [V_{\mu\nu}, A^{\mu\nu}] \ket \,+\, i\lambda^{PVA}_1 \bra P [\nabla_\mu V^{\mu\alpha}, \nabla^\nu A_{\nu\alpha}] \ket\phantom{\frac{1}{2}} \nonumber \\
&&+\, i\lambda^{PVA}_2 \bra P [\nabla^\nu V^{\mu\alpha}, \nabla_\mu A_{\nu\alpha}] \ket
\,+\, i\lambda^{PVA}_3 \bra P [\nabla_\alpha V^{\mu\nu}, \nabla^\alpha A_{\mu\nu}] \ket\phantom{\frac{1}{2}} \nonumber \\
&&+\, i\lambda^{PVA}_4 \bra P [ V^{\mu\nu}, \nabla^2 A_{\mu\nu}] \ket
+\, i\lambda^{PVA}_5 \bra P [ V^{\mu\alpha}, \nabla_\mu \nabla^\nu A_{\nu\alpha}] \ket \phantom{\frac{1}{2}}\nonumber \\
&&+\, i\lambda^{PVA}_6 \bra P \left[ \nabla^\nu \nabla_\mu V^{\mu\alpha},  A_{\nu\alpha}\right] \ket\, ,\phantom{\frac{1}{2}} \\
\Delta \mathcal{L}_{VRR}&=&\,i\,\lambda^{VRR} \bra V^{\mu\nu} \nabla_\mu R \nabla_\nu R \ket \, ,\quad\quad (R=S,P)\phantom{\frac{1}{2}} \\
%
%
\Delta \mathcal{L}_{VVV}&=&\,  i\,\lambda^{VVV}_0 \bra V^{\mu\nu} V_{\mu\alpha} V_\nu^{\,\,\alpha}\ket\,+\,
   i\,\lambda^{VVV}_1 \bra V^{\mu\nu} [\nabla_\mu V_{\alpha\beta},\nabla_\nu V^{\alpha\beta}] \ket  \phantom{\frac{1}{2}}\nonumber \\ &&
 +i\,\lambda^{VVV}_2 \bra V^{\mu\nu} [\nabla^\beta V_{\mu\alpha},\nabla_\beta V_{\nu}^{\,\alpha}] \ket
+i\,\lambda^{VVV}_3 \bra V^{\mu\nu} [\nabla_\mu V_{\beta\alpha},\nabla^\alpha V_{\nu}^{\,\beta}] \ket\phantom{\frac{1}{2}} \nonumber \\ &&
 +i\,\lambda^{VVV}_4 \bra V^{\mu\nu} [\nabla_\mu V_{\nu\alpha},\nabla_\beta V^{\alpha\beta}] \ket  +
i\,\lambda^{VVV}_5 \bra V^{\mu\nu} [\nabla^\alpha V_{\mu\nu},\nabla^\beta V_{\alpha\beta}] \ket\phantom{\frac{1}{2}} \nonumber \\ &&
+i\,\lambda^{VVV}_6 \bra V^{\mu\nu} [\nabla^\alpha V_{\mu\alpha},\nabla^\beta V_{\nu\beta}] \ket
+i\,\lambda^{VVV}_7 \bra V^{\mu\nu} [\nabla^\alpha V_{\mu\beta},\nabla^\beta V_{\nu\alpha}] \ket \phantom{\frac{1}{2}}\,, \hskip 1.2cm\mbox{}\\
%
\Delta \mathcal{L}_{VAA}&=&\,i\,\lambda^{VAA}_0 \bra V^{\mu\nu} A_{\mu\alpha} A_\nu^{\,\,\alpha}\ket
\,+i\,\lambda^{VAA}_1 \bra V^{\mu\nu} [\nabla_\mu A_{\alpha\beta},\nabla_\nu A^{\alpha\beta}] \ket\phantom{\frac{1}{2}} \nonumber \\ &&
+i\,\lambda^{VAA}_2 \bra V^{\mu\nu} [\nabla^\beta A_{\mu\alpha},\nabla_\beta A_{\nu}^{\,\alpha}] \ket
+i\,\lambda^{VAA}_3 \bra \nabla^\beta V^{\mu\nu} [ A_{\mu\alpha},\nabla_\beta A_{\nu}^{\,\alpha}] \ket \phantom{\frac{1}{2}}\nonumber \\ &&
+i\,\lambda^{VAA}_4 \bra V^{\mu\nu} [\nabla_\mu A_{\beta\alpha},\nabla^\alpha A_{\nu}^{\,\beta}] \ket
+i\,\lambda^{VAA}_5 \bra \nabla_\mu V^{\mu\nu} [ A_{\beta\alpha},\nabla^\alpha A_{\nu}^{\,\beta}] \ket \phantom{\frac{1}{2}}\nonumber \\ &&
 +i\,\lambda^{VAA}_6 \bra \nabla^\alpha V^{\mu\nu} [ \nabla_\mu A_{\beta\alpha}, A_{\nu}^{\,\beta}] \ket
+i\,\lambda^{VAA}_7 \bra V^{\mu\nu} [\nabla_\mu A_{\nu\alpha},\nabla_\beta A^{\alpha\beta}] \ket\phantom{\frac{1}{2}} \nonumber \\ &&
+i\,\lambda^{VAA}_8 \bra \nabla_\mu V^{\mu\nu} [ A_{\nu\alpha},\nabla_\beta A^{\alpha\beta}] \ket
+i\,\lambda^{VAA}_9 \bra \nabla_\beta V^{\mu\nu} [\nabla_\mu A_{\nu\alpha}, A^{\alpha\beta}] \ket\phantom{\frac{1}{2}} \nonumber \\ &&
+i\,\lambda^{VAA}_{10} \bra V^{\mu\nu} [\nabla^\alpha A_{\mu\nu},\nabla^\beta A_{\alpha\beta}] \ket
+i\,\lambda^{VAA}_{11} \bra V^{\mu\nu} [\nabla^\alpha A_{\mu\alpha},\nabla^\beta A_{\nu\beta}] \ket\phantom{\frac{1}{2}} \nonumber \\ &&
+i\,\lambda^{VAA}_{12} \bra \nabla^\alpha V^{\mu\nu} [ A_{\mu\alpha},\nabla^\beta A_{\nu\beta}] \ket
+i\,\lambda^{VAA}_{13} \bra V^{\mu\nu} [\nabla^\alpha A_{\mu\beta},\nabla^\beta A_{\nu\alpha}] \ket \phantom{\frac{1}{2}}\nonumber \\ &&
+i\,\lambda^{VAA}_{14} \bra \nabla^\alpha V^{\mu\nu} [ A_{\mu\beta},\nabla^\beta A_{\nu\alpha}] \ket \, .\phantom{\frac{1}{2}}
\end{eqnarray}
All coupling constants are real, $M_R$ are the resonance masses, the brackets $\langle ... \rangle$ denote a trace of the corresponding flavour matrices, and the standard notation defined in Refs.~\cite{RChTa,RChTc} is adopted.

As our Lagrangian $\mathcal{L}_{R\chi T}$ satisfies the $N_C$ counting rules for
an effective theory with $U(3)$ multiplets, only operators that have one trace
in the flavour space are considered~\cite{KL:00}. The different fields,
masses and momenta are of $\cO(1)$ in the $1/N_C$ expansion. Taking into account
the interaction terms, one can check that
$F,\,F_V,\,G_V,\,F_A,\,c_d,\,c_m $ and $d_m$ are of $\cO(\sqrt{N_C})$;
$\lambda_i^{R_1R_2}$ of order $\cO(N_C^0)$ and $\lambda_i^{R_1R_2R_3}$ of order
$\cO(1/\sqrt{N_C})$.
 The mass dimension of these parameters is
$[F]=[F_V]=[G_V]=[F_A]=[c_d]=[c_m]=[d_m]=[\lambda_0^{R_1R_2R_3}]=E$,
$[\lambda_i^{R_1R_2}]=E^0$ and $[\lambda_{i\not=0}^{R_1R_2R_3}]=E^{-1}$.

Note that the $U(3)$ equations of motion have been used in order to reduce the number
of operators. For instance, terms like $\langle P\,\nabla_\mu u^\mu \rangle$ are
not present in Eq.~(\ref{P}), since using the equations of motion they can be transformed
into operators that, either have been already considered, or contain a higher number of resonance fields.

The \rcht\ Lagrangian~\eqn{lagrangian} contains a large number of unknown coupling constants. However, as we will see in the next sections, the short-distance QCD constraints allow to determine many of them.

\section{Form factors and correlators at NLO in $1/N_C$}\label{sec:ff}

Let us consider the two-point correlation function of two currents in the chiral limit:
\begin{eqnarray}
\Pi_{XX}^{\mu\nu}(q)&\equiv& i\int \mathrm{d}^4x \, \mathrm{e}^{iqx}\;
\langle 0|T\left(J_X^\mu(x)J_X^\nu(0)^\dagger\right)|0\rangle \,
=\, \left( -g^{\mu\nu} q^2 + q^\mu q^\nu \right)\,\Pi_{XX}(q^2)\, , \nonumber \\
\Pi_{YY}(q)&\equiv& i\int \mathrm{d}^4x \, \mathrm{e}^{iqx}\;
\langle 0|T\left(J_Y(x)J_Y(0)^\dagger\right)|0\rangle \, , \label{def1}
\end{eqnarray}
where $J_X^\mu(x)$ can denote the vector or axial-vector currents and $J_Y(x)$ the scalar or pseudo-scalar densities,
\begin{equation}
\begin{array}{rlrl}
J^\mu_V &=\, \bar{d}\gamma^\mu u \, ,  \qquad \qquad & J_S &=\, \bar{d}  u \, , \nonumber \\
J^\mu_A &=\, \bar{d}\gamma^\mu \gamma_5 u \, , \qquad \qquad & J_P &=\, i\, \bar{d} \gamma_5  u \, .
\end{array}
\label{def2}
\end{equation}

The associated spectral functions
are a sum of positive contributions corresponding to the different intermediate states.
At large $q^2$, the vector and axial-vector spectral functions tend to a constant
whereas the scalar and pseudo-scalar ones grow like $q^2$~\cite{v-aQCD,Weinberg:1967kj,SVZ,Jamin:1994vr}. 
Therefore, since there is an
infinite number of possible states,
the absorptive contribution in the spin--1 correlators coming from each
intermediate state should vanish in the $q^2 \rightarrow \infty$ limit
if we assume a similar short-distance behaviour for all of them.
The high-energy behaviour of the spin--0 spectral functions
is not so clear as, {\it a priori},
a constant behaviour for each intermediate cut does not seem to be excluded.
However, the fact  that $\Pi_{SS}(t) - \Pi_{PP}(t)$
vanishes as $1/t^2$ in the chiral limit~\cite{SVZ,Jamin:1994vr}, 
the Brodsky-Lepage rules for the form factors~\cite{brodsky-lepage}
and the $1/t$ behaviour of each one-particle intermediate cut (tree-level
exchanges) seems to point out  that
every absorptive contribution to $\mathrm{Im}\Pi_{YY}(t)$
must also vanish at large momentum transfer.

At leading order in $1/N_C$ the two-point correlation functions reduce to tree-level exchanges of meson states with the appropriate quantum numbers. At the next-to-leading order, they get contributions from two-particle exchanges and, therefore, one needs to consider quantum loops involving virtual resonance propagators. The ultraviolet behaviour of these quantum loops was analyzed for the pion vector form-factor and $\Pi_{SS-PP}(t)$ in Refs.~\cite{RSP:05,natxo-tesis,L8-nlo}. We will present here a more general analysis, although focusing in particular in the $\Pi_{VV-AA}(t)$ correlator.

The optical theorem relates the two-particle spectral cuts with the corresponding two-body form factors. A tree-level calculation of the form factors determines the spectral function at the next-to-leading order in $1/N_C$. The complete correlator can then be reconstructed through a dispersive calculation, up to possible subtraction constants \cite{L8-nlo}.

We have calculated all two-body form factors associated with the scalar, pseudoscalar, vector and axial-vector currents, generated by the \rcht\ Lagrangian discussed in the previous section, and have analysed their explicit relations with the spectral functions, studying their ultraviolet behaviour. In the simplest cases with just one form-factor $\mF_{m_1,m_2}(t)$,  the relation takes the form
\begin{equation}
\left. \mathrm{Im}\, \Pi(t)\right|_{m_1,m_2} \quad
= \quad \xi(t)\,\, |\mF_{m_1,m_2}(t)|^2\, ,
\end{equation}
with $\xi(t)$ a known kinematic factor that depends on the considered channel.
Imposing that the spectral function must vanish as $1/t$ at $t \rightarrow \infty$ yields that $\mF_{m_1,m_2}(t)$ has to behave in a given way depending on $\xi(t)$. Thus, some constraints on the effective parameters are found.
In appendix~A, we give the whole list of form factors in the even-intrinsic-parity sector
of R$\chi$T in the SRA,
the exact relations between them and the spectral functions, the
constraints derived from the high-energy analysis, and
the structure of the form factors after imposing the proper short-distance behaviour.
Some of them can be found in former literature~\cite{polychromatic,RSP:05}.

One of our aims is to clarify the status of form factors involving resonances as asymptotic states, how they must behave at short distances and which constraints can be extracted. Although their status as external Fock states can be questionable, the presence of resonance states at intermediate loops is unavoidable if the hadrons are to be described through a quantum field theory. As an implication of this, well behaved amplitudes with resonances as external states should be studied
when considering calculations at the one loop level. We have also analysed form factors involving one photon and one meson in the final state, but no new constraints emerge from their short-distance analysis. Thus, we find that the two-meson form factor analysis provides the most stringent set of constraints.

\begin{figure}
\begin{center}
\includegraphics[scale=0.8]{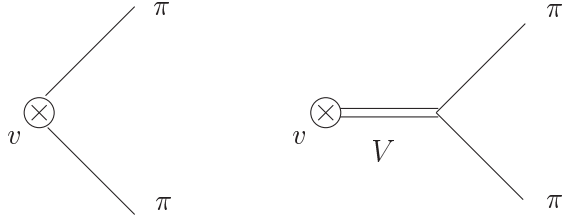}
\caption{\label{vgg}
Tree-level contributions to the vector form factor of the pion.
A single line stands for a pseudo-Goldstone boson while
a double line indicates a vector resonance.}
\end{center}
\end{figure}

As an example, we show here the case of the pion form factor, defined through the two-pseudo-Goldstone matrix
element of the vector current:
\begin{eqnarray}
\bra \pi^0 (p_1) \pi^- (p_2) | \bar{d}\gamma^\mu u | 0\ket &=& \sqrt{2}\, \mathcal{F}_{\pi\pi}^{\,v} (q^2)\, (p_2-p_1)^\mu \, .
\end{eqnarray}
The diagrams contributing at leading order in $1/N_C$ are shown in Figure~\ref{vgg}. They generate the result
\begin{eqnarray}
\mathcal{F}_{\pi\pi}^{\,v} (t)&=&1\,+\,\frac{F_VG_V}{F^2} \frac{t}{M_V^2-t} \, . \label{vffpio}
\end{eqnarray}
By means of the optical theorem, the corresponding imaginary part of the vector correlator is found to be
\begin{eqnarray}
\mathrm{Im}  \Pi_{VV} (t) |_{\pi\pi}  &=&
\frac{\theta(t)}{24\pi}\; |\mathcal{F}_{\pi\pi}^{\,v} (t)|^2 \, .
\end{eqnarray}
Since $\mathrm{Im} \Pi_{VV} (t) |_{\pi\pi}$  should vanish
in the limit $t \rightarrow \infty$, the form factor is also constrained to be zero at infinite momentum transfer.
Therefore, the vector couplings should satisfy
\begin{equation}\label{eq:FG=F2}
F_V\, G_V\, =\, F^2 \, ,
\end{equation}
which implies
\begin{eqnarray}
\mF_{\pi\pi}^{\,v} (t)&=&\frac{M_V^2}{M_V^2-t} \, ,
\end{eqnarray}
as we would have obtained imposing the Brodsky-Lepage behaviour in
Eq.~(\ref{vffpio}).

\section{The correlator $V-A$ in \rcht}\label{sec:v-a}

Let us consider the two-point correlation functions of two vector or
axial-vector currents, in the chiral limit. Of particular interest
is their difference $\Pi(t)\equiv \Pi_{VV}(t)-\Pi_{AA}(t)$, which is
identically zero in QCD perturbation theory. When $t\to\infty$, this
correlator vanishes as $1/t^3$~\cite{v-aQCD,Weinberg:1967kj,SVZ}.

In the large--$N_C$ limit, $\Pi(t)$ has the general form
\begin{equation}
\label{eq:LO}
 \Pi(t) \,=\,  \sum_i\, \left[
 \Frac{ 2\, F_{V_i}^2}{ M_{V_i}^2-t }
 \, -\, \Frac{2\, F_{A_i}^2}{ M_{A_i}^2-t} \right]\, +\, \Frac{2F^2}{t} \, ,
\end{equation}
which involves an infinite number of vector and axial-vector resonance
exchanges. This expression can be easily obtained within R$\chi$T,
with $F_{V_i}$ and $F_{A_i}$ being the relevant meson couplings.

At the NLO in $1/N_C$, $\Pi(t)$ has a contribution from one-particle
exchanges plus one-loop corrections $\Pi(t)|_\rho$ which generate absorptive contributions from
two-particle intermediate states. At this order, the corresponding spectral functions of the
vector and axial-vector correlators take the form:
\beqn
\label{eq:ImSS}
 \frac{1}{\pi}\,\imag\Pi_{V}(t) & =& 2 F_V^{r\, 2} \,\delta(t-M_V^{r\, 2} )\, +\,
 \Frac{n_f}{2} \Frac{\rho_V(t)}{24\pi^2}\, ,\no
\\
  \frac{1}{\pi}\,\imag\Pi_{A}(t) & =& 2 F^2 \,\delta(t)\, +\,
  2 F_A^{r \, 2} \, \delta(t-M_A^{r \, 2})\, +\,
  \Frac{n_f}{2} \Frac{\rho_A(t)}{24\pi^2} \,,
\eeqn
with
\beqn
 \rho_V(t) & =& \theta(t)\,  |\mF^v_{\pi\pi}(t)|^2 \,+\,
\theta(t-M_A^2) \left(1-\frac{M_A^2}{t}\right) \!\Bigg\{ \!\!\left( \Frac{M_A^2}{t}\!+4+\!\Frac{t}{M_A^2} \right)\! |\mathcal{F}^{\,v}_{A\pi}|^2 \nonumber \\
&&   +\! \left(1-\frac{M_A^2}{t}\right)^2  \left( \Frac{t}{M_A^2} + \Frac{t^2}{2M_A^4}\right) |\mathcal{G}^{\,v}_{A\pi}|^2
+2 \left(1-\frac{M_A^2}{t}\right) \left( 1+\Frac{2t}{M_A^2} \right) \nonumber \\
&&  \times \mathrm{Re} \{ \mathcal{F}^{\,v}_{A\pi} {\mathcal{G}^{\,v}_{A\pi}}^* \}  \Bigg\} \, +\,\theta(t-M_P^2) \, \left(1-\frac{M_P^2}{t}\right)^3 \frac{t^2}{2}   |\mathcal{F}^{\,v}_{P\pi}|^2
 \,+ \cdots
\\
 \rho_A(t) & =& \theta(t-M_V^2) \left(1-\frac{M_V^2}{t}\right) \Bigg\{ \left( \Frac{M_V^2}{t}+4+\Frac{t}{M_V^2} \right) |\mathcal{F}^{\,a}_{V\pi}|^2
+ \left(1-\frac{M_V^2}{t}\right)^2  \nonumber \\
&&  \times \left( \Frac{t}{M_V^2} + \Frac{t^2}{2M_V^4}\right) |\mathcal{G}^{\,a}_{V\pi}|^2  +2 \left(1-\frac{M_V^2}{t}\right) \left( 1+\Frac{2t}{M_V^2} \right)\mathrm{Re} \{ \mathcal{F}^{\,a}_{V\pi} {\mathcal{G}^{\,a}_{V\pi}}^* \}  \Bigg\}
 \no\\ && +\,\theta(t-M_S^2) \left(1-\frac{M_S^2}{t}\right)^3 |\mathcal{F}^{\,a}_{S\pi} (t)|^2
 \,+ \cdots
\eeqn
The dots stand for contributions with higher thresholds.
Here we just show the lowest-mass two-particle
exchanges: two Goldstone bosons or one Goldstone and one resonance.
In the energy region we are interested in, exchanges of
two heavy resonances are kinematically
suppressed (appendix~C)~\cite{natxo-tesis}. Our normalization takes into account the different
flavour-exchange possibilities.

Using dispersion relations (appendix~C), it is possible to prove that at NLO in $1/N_C$ the correlator $\Pi(t)$
has the structure
\begin{equation}\label{eq:Pi_structure}
\Pi(t)\, =\,
\Frac{2\, F^2}{t} \, + \, \sum_i \left[ \frac{2\, F_{V_i}^{r\,\, 2}}{M_{V_i}^{r\,\,2} \, -\, t} \,
- \, \frac{2\, F_{A_i}^{r\,\,2}}{M_{A_i}^{r\,\, 2} \, -\, t} \right]
\, + \,\sum_{m_1,m_2}  \Pi(t)|_{m_1,m_2}\, ,
\end{equation}
where $\Pi(t)|_{m_1,m_2}$ denote the contributions associated with the two-meson absorptive cuts $m_1,m_2$.
Their imaginary parts are related to the corresponding two-meson form-factors
through the optical theorem (the precise relations are given in appendix~A).
Since $\Pi(t)|_{m_1,m_2}$ should vanish at infinite momentum transfer, the full $\Pi(t)|_{m_1,m_2}$
contribution can be reconstructed from its absorptive part through an unsubtracted dispersion relation. The fact that the form-factors are well behaved at infinite momentum guarantees that the dispersive integrals are convergent.
Notice that analytic polynomial contributions cannot be present in \eqn{eq:Pi_structure} because they would violate the short-distance QCD constraints.

\subsection{Single resonance approximation}

Let us adopt the SRA as our starting point. At leading order in $1/N_C$, $\Pi(t)$ takes then the simpler form:
\begin{equation}
 \Pi(t) \,=\,\Frac{ 2\, F_{V}^2}{ M_{V}^2-t }
 \, -\, \Frac{2\, F_{A}^2}{ M_{A}^2-t} \, +\, \Frac{2F^2}{t}  \, .
\end{equation}
The high-energy behaviour required by the OPE, $\Pi(t)\sim 1/t^3$, implies the first and
second Weinberg sum rules (WSRs)~\cite{v-aQCD,Weinberg:1967kj,SVZ}:
%
%
\begin{equation}
\label{eq:WSR-LO}
 F_{V}^2 \,-\, F_{A}^2 \,=\, F^2 \, ,
\quad \qquad  \qquad
F_{V}^2 \,M_{V}^2\, -\, F_{A}^2 \,M_{A}^2  \,=\,0 \,.
\end{equation}
These relations determine the vector and axial-vector couplings
in terms of the resonance masses:
\begin{equation}
\label{eq:FV_FA}
F_V^2\, =  \, F^2\, \Frac{M_A^2}{M_A^2-M_V^2} \, ,
\quad \quad \qquad F_A^2\,=\,F^2\, \Frac{M_V^2}{M_A^2-M_V^2}\, ,
\end{equation}
with $M_A > M_V$. Using the constraint \eqn{eq:FG=F2}, obtained from the pion form factor,
one gets the additional relation
\begin{eqnarray}
\label{eq:GV}
G_V^2\,=\,F^2 \, \frac{M_A^2-M_V^2}{M_A^2}\, .
\end{eqnarray}

The short-distance behaviour of the $\Pi_{SS-PP}$ correlator \cite{sspp2} and the pion scalar form-factor~\cite{oller} generate similar expressions for the scalar couplings \cite{polychromatic,L8-nlo}
\begin{eqnarray}
c_m^2\, =  \, \frac{F^2}{8}\,   \frac{M_P^2}{M_P^2-M_S^2}\, , \quad \quad && \qquad
d_m^2\,=\,\frac{F^2}{8}\, \frac{M_S^2}{M_P^2-M_S^2}\, ,
\nonumber \\[7pt]
c_d^2\,=\,\frac{F^2}{2}\,\frac{M_P^2-M_S^2}{M_P^2}\, ,
\quad \quad &&\qquad\qquad M_P\, > \, M_S\, .
\label{eq:cd}
\end{eqnarray}
Thus, at LO in $1/N_C$, the couplings $F_V$, $F_A$, $G_V$, $c_m$, $c_d$ and $d_m$ are fixed in terms of
$F$, $M_V, \, M_A, \, M_S$ and $M_P$, the chiral limit values of the pion decay constant and the resonance masses.

We can study in a similar way the form factors needed to compute the two-particle-exchange contributions $\Pi(t)|_{m_1,m_2}$. For the lightest channels, the separate short-distance analysis of each form-factor
allows their determination in terms of the resonance masses and the couplings $F_V$, $G_V$, $F_A$ and $c_d$.
Using the large--$N_C$ relations in Eqs.~(\ref{eq:FV_FA}),~(\ref{eq:GV}) and~(\ref{eq:cd}), the results can be further simplified, leaving the form-factors expressed in terms
of just the resonance masses~\cite{natxo-tesis,L8-nlo}:
\beqn\label{eq:VFF}
\mF_{\pi\pi}^v(t)& =& \frac{M_V^2}{M_V^2-t}\, ,
\no\\
\mF_{A\pi}^v(t)& =&
-\, \sqrt{\frac{M_A^2}{M_V^2}-1}\,\frac{M_V^2}{M_V^2-t}\, ,
\qquad \qquad
\mG_{A\pi}^v(t) \, =\, 0\, ,
\\
\mF_{P\pi}^v(t)& =& 0\, ,\phantom{\frac{1}{2}}
\no\eeqn
\beqn\label{eq:AFF}
\mF_{V\pi}^a(t)& =& -\,\sqrt{1-\frac{M_V^2}{M_A^2}}\,
\frac{M_A^2 (2 M_V^2-t)}{t (M_A^2-t)}\, ,
\qquad \qquad
\mG_{V\pi}^a(t)\, =\,  -\,\sqrt{1-\frac{M_V^2}{M_A^2}}\,
\frac{2 M_A^2 M_V^2}{t (M_A^2-t)}\, ,
\no\\
\mF_{S\pi}^a(t)& =& \,\sqrt{1-\frac{M_S^2}{M_P^2}}\,
\frac{\sqrt{2}\,M_A^2}{M_A^2-t}\, .
\eeqn
The explicit results in terms of \rcht\ couplings can be found in appendix~A and in Ref.~\cite{natxo-tesis}.
These form-factors have been determined by imposing a good high-energy behaviour on the corresponding spectral functions, {\it i.e.} by demanding that the contributions from each absorptive channel to $\rho_V(t)$ and $\rho_A(t)$ should vanish at infinite momentum transfer.
In this first analysis, performed within the SRA,
we have just focused our attention on each separate channel and
we have not used information from other absorptive two-meson cuts
to further simplify  the form-factors.~\footnote{
In Ref.~\cite{Pade}, it has been argued that large discrepancies
may occur between the values of the masses and couplings
of the full large--$N_C$ theory and those from descriptions with a finite number
of resonances. Even in this case, it is found that one can obtain safe
determinations of the LECs as far as one is able to construct a good
interpolator that reproduces the right asymptotic behaviour at  low and high energies.
Further issues related to the truncation of the spectrum to a finite number of resonances are
discussed in Refs.~\cite{inf-res,ximo}.}

As it was found in the case of the scalar and $\Pi_{SS-PP}$ correlators~\cite{natxo-tesis,L8-nlo},
it is quite remarkable that these short-distance constraints completely determine the form-factors in terms of the resonance masses. The form factors $\mF_{P\pi}^v(t)$ and $\mathcal{G}^v_{A\pi}(t)$ turn out to be identically zero, within the SRA.

Once the form factors have been determined, the corresponding contributions to the two-point correlation function can be obtained in a rather straightforward way. The first two-meson contribution to $\Pi(t)$ is given by the $\pi\pi$--cut:
\begin{eqnarray}
 \Pi(t)|_{\pi\pi} \, &=&\,\,\, \Frac{n_f}{2}
 \Frac{1}{24\, \pi^2} \, \left(\Frac{M_V^2}{M_V^2\, -\, t}\right)^2 \,
 \left[ -1 + \Frac{t}{M_V^2} - \ln{\Frac{-t}{M_V^2}}
\right] \,. \label{totc1}
\end{eqnarray}
The double pole at $t=M_V^2$ is generated by the large--$N_C$ expression for the pion form factor in Eq.~\eqn{eq:VFF}.

Away from the resonance peak, where the perturbative $1/N_C$ expansion is valid,
the vector meson width generates the double pole structure. This can be easily realized
by rewriting the sum of tree-level and the $\Pi(t)|_{\pi\pi}$ contribution in the form
\begin{equation}
\label{eq.Gamma}
\Pi(t) \, =\, \Frac{ 2 F_V^{r\,\, 2}}{M_V^{r\,\, 2}-t}
\, \left\{  \, 1\, \,
- \, \Frac{1}{\pi} \Frac{\Gamma_{_{V\to\pi\pi}}}{M_V}
- \, \Frac{1}{\pi} \Frac{M_V \Gamma_{_{V\to\pi\pi}}}{M_V^2-t}\, \ln{\Frac{-t}{M_V^2}}
\,\, +\, ...\, \right\} \, ,
\end{equation}
where the dots stand for higher channels and  the $V$--meson decay width into two Goldstones is given
in the chiral limit by
$\Gamma_{_{V\to\pi\pi}}=\frac{n_f}{2} \frac{G_V^2 M_V^3}{48\pi F^4}$~\cite{Dyson-VFF}.
Likewise, this expression shows in a manifest way how the formal $1/N_C$ suppression
works in terms of the physical hadronic parameters.

The next (ordered by threshold) absorptive cuts correspond to $V\pi$, $S\pi$, $A\pi$ and $P\pi$.
Due to the complexity of the results their precise expressions are relegated to appendix~B.
It is possible to show that states with higher energy thresholds
turn out to be more and more suppressed (see appendix C.2). Thus, only the contributions
from cuts with at most one resonance field have been taken into account:
$\pi\pi$, $P\pi$ and $A\pi$ for the vector correlator and $V\pi$ and $S\pi$ for the axial-vector one.
All the results for particular channels from appendix~A, obtained in the two-flavour case,
have been multiplied by a factor $n_f/2$ in order to give the general result for $n_f$ light flavours
shown in appendix~B.

At large values of $t$, the one-loop contribution has the behaviour
\begin{equation}
\left.\Pi(t)\right|_{\rho} \,=\,  \frac{2F^2}{t}\,
\left( \delta_{_{\rm NLO}}^{(1)} +  \widetilde{\delta}_{_{\rm NLO}}^{(1)} \ln{\Frac{-t}{M_V^2}}  \right)
\, +\,
\frac{2F^2 M_V^2}{t^2} \,\left( \delta_{_{\rm
NLO}}^{(2)}\, +\, \widetilde{\delta}_{_{\rm NLO}}^{(2)} \,
\ln\frac{-t}{M_V^2}\right)\, +\, \cO\left(\frac{1}{t^3}\right)\, .
\end{equation}
Since the logarithmic terms $\ln(-t)/t$ and $\ln(-t)/t^2$ should vanish, one obtains
the constraints:
\begin{equation}
\widetilde{\delta}_{_{\rm NLO}}^{(1)}=\widetilde{\delta}_{_{\rm NLO}}^{(2)}=0\, .\label{masses}
\end{equation}
Taking into account the tree-level contributions and imposing
the right short-distance behaviour, $\Pi(t)\sim 1/t^3$, one gets the additional relations:
\begin{eqnarray}
F^2\, (1\,+\,\delta_{_{\rm NLO}}^{(1)})\, - \, F_V^{r\, 2} \, +\, F_A^{r\, 2}\, = \, 0 \, ,
\no\\[7pt]
F^2 \,M_V^2 \,\delta_{_{\rm NLO}}^{(2)}\, - \, F_V^{r\, 2}\, M_V^{r\, 2} \,
+\, F_A^{r\, 2}\, M_A^{r\, 2} \, = \,0 \, , \label{eq:NLO_rel}
\end{eqnarray}
which determine the effective couplings $F_V^r$ and $F_A^r$ up to NLO in $1/N_C$:
\begin{eqnarray}
F_V^{r\,\,2}&=& F^2 \frac{M_A^{r\,\,2}}{M_A^{r\,\,2}-M_V^{r\,\,2}}
        \left(1+\delta_{_{\rm NLO}}^{(1)}-\frac{M_V^2}{M_A^2}\delta_{_{\rm NLO}}^{(2)}  \right)
\, ,\nn \\[7pt]
F_A^{r\,\,2}&=& F^2 \frac{M_V^{r\,\,2}}{M_A^{r\,\,2}-M_V^{r\,\,2}}
        \left(1+\delta_{_{\rm NLO}}^{(1)}-\delta_{_{\rm NLO}}^{(2)} \right)
\, .\label{dmr}
\end{eqnarray}

Within the SRA, the conditions \eqn{masses} have the unique solution, $M_A=M_V$, $M_P=\sqrt{2} M_S$.
Note that the whole $A\pi$ and $V\pi$ contributions to $\Pi(t)$ are then identically zero, while $\Pi(t)|_{S\pi}$ cancels the leading high-energy behaviour of $\Pi(t)|_{\pi\pi}$. The $\delta^{(m)}_{_{\rm NLO}}$ corrections are given by
\begin{eqnarray}
\delta_{_{\mathrm{NLO}}}^{(1)} &=&
\Frac{n_f}{2} \Frac{M_V^2}{48\pi^2F^2}
\left\{ 1
-  \Frac{M_S^{2}}{M_V^{2}}
\left[
3 \left(1-\Frac{M_S^2}{M_V^2}\right)^2 \ln{\left(1\,-\, \Frac{M_V^2}{M_S^2}\right)}
+\Frac{3M_S^2}{M_V^2} -\Frac{9}{2} +\Frac{M_V^2}{M_S^2}
\right]
\right\}  \,,
\nn\\[7pt]
\delta_{_{\mathrm{NLO}}}^{(2)} &=&
\Frac{n_f}{2} \Frac{M_V^2}{48\pi^2F^2}
\left\{ 1
-  \Frac{M_S^{4}}{M_V^{4}}
\left[ \left( \Frac{2M_S^2}{M_V^2} -3 +\Frac{M_V^4}{M_S^4}\right)
 \ln{\left(1\,-\, \frac{M_V^2}{M_S^2}\right)}
 \, -\, \Frac{M_V^4}{M_S^4}\ln{\Frac{M_V^2}{M_S^2}}
\right.\right.\nn
\\
&&\qquad\qquad\qquad\qquad\quad\left.\left.
 \,+\, 2-\Frac{2 M_V^2}{M_S^2}+\Frac{M_V^4}{M_S^4}
\right]\right\}  \,.
\\ \nn
\end{eqnarray}

\subsection{Numerical impact of heavier vector and axial multiplets}

The Resonance Chiral Theory is an effective approximation to QCD that models
large-$N_C$ by truncating the tower of resonances to a finite number. However, an infinite
number of resonances is needed to recover the correct QCD behaviour~\cite{polychromatic,NC,inf-res}.
Therefore, it should not be surprising to find conflicts between the short-distance constraints
as one analyzes a wider and wider set of QCD matrix elements.
This inconsistence between constraints has popped up in previous analysis of
three-point Green-functions~\cite{VAP,ximo} and it also arises
when comparing the short-distance constraints from vector, axial-vector, scalar and
pseudo-scalar form-factors within the SRA (appendix~A)~\cite{natxo-tesis}.


These incompatibilities can always be solved by including additional resonance multiplets.
We follow the Minimal Hadronic Approximation~\cite{MHA}, and only include a minimal set of resonance multiplets such that all relevant short-distance constraints are satisfied for the problem at hand.

Our explicit form factor expressions in Eqs.~\eqn{eq:VFF} and \eqn{eq:AFF} have been obtained analysing each form factor separately. However the assumed set of short-distance constraints is not fully consistent. The results compiled in
appendix~A show the existence of two inconsistent conditions: the restrictions for $\lambda_i^{VA}$ in Table \ref{tableVFF} from the vector form factor to an axial resonance field and a pion, and those
in Table \ref{tableAFF} from the axial form factor to a vector resonance field and a
pion are incompatible. The simpler solution is the inclusion of a second
multiplet of vector ($V'$) and axial-vector ($A'$) resonances. One gets in this way a consistent set of constraints for the couplings of the lowest mass multiplets $S$, $P$, $V$ and $A$; any possible problem is then relegated to the heavier states, that produce mild effects on the region of validity of our effective description
(of course, if one was interested in physical form factors involving $V'$ or $A'$ as external states, the addition of even heavier multiplets would push the problem to the next level in the resonance towers).

In practice, one adds to the \rcht\ Lagrangian the necessary pieces involving the new multiplets
$V'$ and $A'$.
%
%
%
%
%
The reanalysis of the $A\pi$ vector form-factor and the $V\pi$ axial form factor of appendix~A
yields, respectively, the constraints
\begin{eqnarray}
& F_V(2\lambda^{VA}_2-2\lambda^{VA}_3+\lambda^{VA}_4+2\lambda^{VA}_5)+
F_V'(2\lambda^{V'A}_2-2\lambda^{V'A}_3+\lambda^{V'A}_4+2\lambda^{V'A}_5)=F_A\,,
& \nonumber \\[7pt]
& F_V(-2\lambda^{VA}_2+\lambda^{VA}_3)+
F_V'(-2\lambda^{V'A}_2+\lambda^{V'A}_3) = 0 \, ,
&
\end{eqnarray}
\begin{eqnarray}
& F_A(2\lambda^{VA}_2-\lambda^{VA}_4-2\lambda^{VA}_5)+
F_A'(2\lambda^{VA'}_2-\lambda^{VA'}_4-2\lambda^{VA'}_5)=-F_V+2G_V   ,
& \nn \\[7pt]
& F_A(-2\lambda^{VA}_2+\lambda^{VA}_3)+
F_A'(-2\lambda^{VA'}_2+\lambda^{VA'}_3)=-G_V \, ,
&
\end{eqnarray}
so the incompatibility is not present any longer.


Once these second multiplets are considered, the large--$N_C$ constraints for $F_V,\, F_A,\, G_V$
are obviously modified (the couplings of any effective Lagrangian contain the information on the heavier states not included in the effective theory). Eqs.~\eqn{eq:FV_FA} and \eqn{eq:GV} take now the form:
\begin{eqnarray}
F_V^2 &=& F^2\, \Frac{M_A^2}{M_A^2-M_V^2}
\, \left(1+\epsilon_{1}-\Frac{M_V^2}{M_A^2} \epsilon_{2}\right) \, ,
\nn \\[7pt]
 F_A^2 &=& F^2\, \Frac{M_V^2}{M_A^2-M_V^2}
\, \left(1+\epsilon_{1} -\epsilon_{2}\right)\, ,
\nn\\[7pt]
G_V^2 &=& F^2 \, \frac{M_A^2-M_V^2}{M_A^2}
\, \Frac{ \left(1-\epsilon_{3}\right)^2}{ 1+\epsilon_{1}-\frac{M_V^2}{M_A^2}\epsilon_{2}}
\, ,
\label{eq:GVbis}
\end{eqnarray}
with the corrections $\epsilon_{i}$ given by
\begin{equation}
\epsilon_{1}=\Frac{F_{A'}^2}{F^2} -\Frac{F_{V'}^2}{F^2} ,
\qquad
\epsilon_{2}=\Frac{F_{A'}^2 M_{A'}^2-F_{V'}^2 M_{V'}^2 }{F^2 M_V^2},
\qquad
\epsilon_{3}=\Frac{F_{V'} G_{V'}}{F^2}\, . \label{FVprime}
\end{equation}
The corrections $\epsilon_{i}$ seem to produce a tiny effect. The value
of $\epsilon_{3}\approx 0.007$ was extracted from
the analysis of the ALEPH data on the pion vector form factor~\cite{Dyson-VFF}.
On the other hand, the assumed convergence of the Weinberg sum-rules
and its phenomenological success~\cite{polychromatic}
seems to point out that $|\epsilon_{1}|,|\epsilon_{2}|\ll 1$.
For this reason, in our numerical calculations we will only take the $\epsilon_{i}$ corrections into account when they appear at LO in $1/N_C$. They will be neglected whenever they enter into contributions which are NLO in $1/N_C$,
as they are a correction to a correction.
Notice that we have introduced both a vector and an axial-vector multiplet
in order to keep the assumed convergence of the $V-A$ Weinberg sum-rules.
Considering the $V'$ contribution alone would lead very likely to large corrections
$\epsilon_{1},\epsilon_{2}\sim 1$ and a lost of the convergence in the sum-rule.

The calculation of the seven form factors we are interested in is straightforward.
The only novelty is the introduction of $V'$ and $A'$.
Due to their heavier thresholds, we neglect any two-particle channels including these higher multiplets
(appendix~C). Only the single-resonance exchange contribution has been considered for the new $V'$ and $A'$ states.
Since we have now a much larger set of couplings, the Brodsky-Lepage form factor constraints~\cite{brodsky-lepage} and the OPE asymptotic behaviour of the two-point Green-functions are no longer enough to fully determine the form-factors. One needs to fix some combinations of the $\lambda_i^{V^{(\prime)}A^{(\prime)}}$ couplings by other means.
Fortunately,
we can use the known constraints coming from the $<\!\!VAP\!\!>$ Green-function analysis of Ref.~\cite{VAP}. The information from the $\bra VAP\ket$ correlator was combined in Ref.~\cite{VAP} with the two Weinberg sum rules and a vanishing high-energy behaviour
for $\mathcal{F}^a_{\pi\gamma}$ and $\mathcal{F}^v_{\pi\pi}$.~\footnote{The spirit of Ref.~\cite{VAP} is to consider form factors with stable states (pseudo-Goldstone modes and on-shell photons).
}
Considering that Ref. \cite{VAP} only used the lowest-lying resonance multiplets, their constraints are right up to $\cO(\epsilon)$ corrections,
which we assume to be tiny:
\begin{eqnarray}
 2\lambda^{VA}_2-\lambda^{VA}_4-2\lambda^{VA}_5 \,=\,-\frac{F_V}{F_A}+\frac{2G_V}{F_A}
 \quad +\quad \cO(\epsilon) \,  ,
 \nn\\
-2\lambda^{VA}_2+\lambda^{VA}_3\,=\,-\frac{G_V}{F_A}
\quad + \quad \cO(\epsilon) \, . \label{VAP}
\end{eqnarray}
Note that it is not a surprise that these constraints are equivalent to the ones coming from the axial form factor to a vector resonance and a pion (Table~\ref{tableAFF}), because this form factor is related to the axial form factor to a photon and a pion, considered in Ref.~\cite{VAP}.
Taking into consideration that the couplings $\lambda_i^{VA}$ appear in $\Pi(t)$ only at NLO in $1/N_C$, we will neglect these $\cO(\epsilon)$ terms.

Using Eqs.~(\ref{eq:GVbis}), the $\bra VAP\ket$ constraint
and imposing the right short-distance behaviour, one determines the new form factors:
\begin{eqnarray}
\mF_{\pi\pi}^{\,v} (t)&=&  \frac{M_V^2}{M_V^2-t}
\quad + \quad \cO(\epsilon) \, ,
\nn \\
\mF^{\,v}_{A\pi} (t) &=&
\left(\frac{M_A^2}{M_V^2}-1\right)^{-\frac{1}{2}}
\, \, \left[
\Frac{M_{V'}^2-M_A^2}{M_{V'}^2-t}- \frac{M_A^2}{M_V^2} \Frac{(M_A^2-t)(M_{V'}^2-M_V^2)}{(M_V^2-t)(M_{V'}^2-t)}
\right]
\quad + \quad \cO(\epsilon) \,, \nonumber \\
\mG^{\,v}_{A\pi} (t)&=& -\, \sqrt{\Frac{M_A^2}{M_V^2}-1}\,\,
\frac{2 M_A^2 \, (M_{V'}^2-M_V^2)}{(M_V^2-t)(M_{V'}^2-t)}
\quad +\quad \cO(\epsilon) \, ,
\nn \\
\mF^{\,v}_{P\pi} (t) &=&
\sqrt{\Frac{M_P^2}{M_S^2}-1}
\, \, \Frac{2 \, (M_{V'}^2-M_V^2)}{(M_V^2-t)(M_{V'}^2-t)}
\quad +\quad \cO(\epsilon)
\,,
\end{eqnarray}
\begin{eqnarray}
\mF^{\,a}_{V\pi} (t) &=&
-\, \sqrt{1-\Frac{M_V^2}{M_A^2}} \,\,
\Frac{M_A^2 \, (2 M_V^2 -t)}{t\, (M_A^2-t)}
\quad +\quad  \cO(\epsilon) \, ,
\nonumber \\
\mG^{\,a}_{V\pi} (t) &=&
-\, \sqrt{1-\Frac{M_V^2}{M_A^2}}\,\, \Frac{2 \, M_V^2\,  M_A^2}{t\, (M_A^2-t)}
\quad +\quad \cO(\epsilon) \, ,
\nn\\
\mF^{\,a}_{S\pi} (t) &=&\sqrt{1-\Frac{M_S^2}{M_P^2}}
\,\, \Frac{\sqrt{2}M_{A'}^2}{M_{A'}^2-t}
\quad +\quad \cO(\epsilon)\, .
\end{eqnarray}
In order to get more compact expressions, this time we have used
the information of some form-factors to simplify others.
Due to the consideration of a higher number of resonance multiplets the
inconsistences between channels have disappeared, in agreement with the assumed convergence
to the full set of large--$N_C$ relations as more and more states are progressively included in the theory.
The $\mF^v_{P\pi}$ form factor has been simplified using the short-distance $\mF^p_{V\pi}$ constraint
$ \lambda^{PV}_1 = \sqrt{2} G_V/ 4d_m$  (appendix~A).
In $\mF^a_{S\pi}$ we have used the relation extracted from $\mF^s_{A\pi}$,  $\lambda_1^{SA}=0$.
Finally, the constraints in Eq.~(\ref{VAP}) have been employed
to simplify the vector--$A\pi$ and axial--$V\pi$ form factors.
Since the $\cO(\epsilon)$ corrections are neglected in the NLO terms,
$\Pi(t)|_\rho$ is known in terms of just the resonance masses and $F$.
Notice that, up to $\cO(\epsilon)$ corrections,
the $\pi\pi$ and $V\pi$ form factors remain the same as in the SRA
and the only change in the axial $S\pi$  form factor is the replacement $M_A\to M_{A'}$.

It would be possible to add as well an extra multiplet of scalar and pseudo-scalar resonances,
$S'$ and $P'$. However, these mesons can only appear in the
$V-A$ correlator within loops, never at tree-level. Hence,
their contributions will be suppressed due to their high threshold.

When the $V'$ and $A'$ resonances are included in the analysis
the resulting expression for $\Pi(t)$ becomes much more complex, though the formal structure
remains exactly the same.
The conditions $\widetilde{\delta}_{_{\rm NLO}}^{(1)}=\widetilde{\delta}_{_{\rm NLO}}^{(2)}=0$
allow to determine $M_{A'}$ and $M_{V'}$:
\begin{eqnarray}
M_{V'}^2 &=&  M_A^2  \, ,
\nonumber \\
M_{A'}^4 &=& \Frac{1}{2} \Frac{M_V^4}{1-\frac{M_S^2}{M_P^2}}
  \left[
  1
  + \Frac{M_A^4}{M_V^4}\left(1-\Frac{M_V^2}{M_A^2}\right)\left(7-\Frac{2 M_A^2}{M_V^2}\right)
  + 2 \left(\Frac{M_P^2}{M_S^2}-1\right)  \left(1-\Frac{M_{V'}^2}{M_V^2}\right)^2
  \right.
\nonumber \\
& & \qquad\qquad\quad
  \left.
  + \left(1-\Frac{M_V^2}{M_A^2}\right) \left(\Frac{2M_{V'}^6}{M_V^6} - \Frac{5 M_A^2 M_{V'}^4}{M_V^6}
  + \Frac{6 M_A^2 M_{V'}^2}{M_V^4} +\Frac{2M_A^2}{M_V^2}\right)
  \right]
 . \label{MVpMAp}
\end{eqnarray}
The new NLO corrections $\delta^{(m)}_{_{\rm NLO}}$ are
given by
\begin{eqnarray}
\delta_{_{\mathrm{NLO}}}^{(m)} &=&
\Frac{n_f}{2} \Frac{M_V^2}{48\pi^2F^2}
\left\{ 1+ \!\left(1-\Frac{M_V^2}{M_A^2}\right)\! \Frac{M_V^2}{M_{V'}^2-M_V^2} \xi_{A\pi}^{(m)}
+2\!\left(\Frac{M_P^2}{M_S^2}-1\right)\! \Frac{M_V^2}{M_{V'}^2-M_V^2} \xi_{P\pi}^{(m)} \right.\nonumber \\
&& \qquad \qquad \qquad \qquad \left.
-  \left(1-\Frac{M_V^2}{M_A^2}\right) \xi_{V\pi}^{(m)}
-  \Frac{2 M_S^{2m}}{M_V^{2m}}
   \left(1-\Frac{M_S^2}{M_P^2}\right) \xi_{S\pi}^{(m)}
\right\}  \,,
\end{eqnarray}
which are known functions of the resonance masses. The different contributions to $\Pi(t)$ and the coefficient functions $\xi_{m_1m_2}^{(m)}$ are relegated to appendix~B.

Note that in this case the determination of $F_V^r$ and $F_A^r$ changes slightly. One can easily reanalyse Eqs.~(\ref{dmr}) to find that now
\begin{eqnarray}
F_V^{r\,\,2}&=& F^2 \frac{M_A^{r\,\,2}}{M_A^{r\,\,2}-M_V^{r\,\,2}}\;
        \left[1+\epsilon_{1}+\delta_{_{\rm NLO}}^{(1)}
        -\frac{M_V^2}{M_A^2} \left(\epsilon_{2}+  \delta_{_{\rm NLO}}^{(2)}\right) \right]
\, ,\nn \\[7pt]
F_A^{r\,\,2}&=& F^2 \frac{M_V^{r\,\,2}}{M_A^{r\,\,2}-M_V^{r\,\,2}}\;
        \left[1+\epsilon_{1}+\delta_{_{\rm NLO}}^{(1)}-\epsilon_{2}-\delta_{_{\rm NLO}}^{(2)} \right]
\, ,\label{dmrbis}
\end{eqnarray}
where we have approximated $F_{V'}^r\simeq F_{V'}$ , $F_{A'}^r\simeq F_{A'}$, since
the effect of the second multiplet is expected to be numerically small and the difference between the LO and the NLO couplings would represent a subleading correction to an already tiny contribution.
%

\section{The chiral couplings $L_{10}^r(\mu)$ and $C_{87}^r(\mu)$}\label{sec:L10}

The low-momentum expansion of $\Pi(t)$ is
determined by \chpt\ \cite{ChPTp4,ChPTp6,op6-correlator}:
\begin{eqnarray}
  \Pi(t) &=&  \frac{2 F^2}{t} \, - \, 8 L_{10}^r(\mu)
\, -\, \frac{\Gamma_{10}}{4\pi^2} \left( \Frac{5}{3}-\ln \frac{-t}{\mu^2} \right)\no
\\
&&
\quad +\,
\frac{t}{F^2} \left[ 16\,C_{87}^r (\mu)  -\frac{\Gamma_{87}^{(L)}}{2\pi^2}
\left( \Frac{5}{3}-\ln \frac{-t}{\mu^2} \right) +\cO\left(N_C^{0}\right) \right]\,+\,
 \cO\left(t^2\right)
 \, ,\label{eq:Pi_chpt}
\end{eqnarray}
with $\Gamma_{10} = -1/4$~\cite{ChPTp4} and  $\Gamma_{87}^{(L)} = - L_9/2$~\cite{ChPTp6}. The couplings $F^2$, $L_{10}$ and $C_{87}/F^2$ are of $\cO(N_C)$, while $\Gamma_{10}$ and $\Gamma_{87}^{(L)}/F^2$ are of $\cO(N_C^0)$ and represent a NLO effect.

On the other hand, the low-energy expansion of \eqn{eq:LO}
determines the chiral LECs at large $N_C$~\cite{RChTa,RChTc,VAP}:
\begin{eqnarray}
L_{10}  &=& \, -\, \Frac{F_V^2}{4 M_V^2} \, +\, \Frac{F_A^2}{4 M_A^2}
 \,= \, -\, \Frac{F^2}{4}\, \left(\Frac{1}{M_V^2}+\Frac{1}{M_A^2}\right) \, .\no \\
C_{87}&=&   \Frac{F^2 F_V^2}{8 M_V^4} \, -\, \Frac{F^2 F_A^2}{8 M_A^4}
 \,= \, \, \Frac{F^4}{8}\,
 \left(\Frac{1}{M_V^4}+\Frac{1}{M_V^2 M_A^2} +\Frac{1}{M_A^4}\right)
\, ,
\label{eq:L10-LO}
\end{eqnarray}
where we have used the relations in Eq.~\eqn{eq:FV_FA} in order to simplify the final
results. Using $M_V\simeq 0.77$~GeV~\cite{PDG} and $M_A\simeq 1$~GeV~\cite{jorge_vicent},
one gets the large--$N_C$ estimates
$L_{10} \approx -5.3 \cdot 10^{-3}$ and $C_{87}\approx 4.3 \cdot 10^{-5}$.

At $\mu_0=770$~MeV,  the phenomenological determination
$L_{10}^r(\mu_0)= (-5.5\pm 0.7)\cdot 10^{-3}$~\cite{polychromatic,L10-Bijnens}
agrees very well with the large--$N_C$ estimate. A slightly smaller absolute value,
$L_{10}^r(\mu_0)= (-5.13\pm 0.19)\cdot 10^{-3}$, was obtained from a fit to the ALEPH
$\tau$ decay data \cite{DGHS:98}.
The large--$N_C$ result for $C_{87}$ is also in good
agreement with the value $C_{87}=(4.5\pm 0.4)\cdot 10^{-5}$, obtained recently
from a series of Pade approximants to large--$N_C$ QCD~\cite{C87-Peris}, using as input the measured resonance spectrum.

Large--$N_C$ estimates like those in Eqs.~(\ref{eq:L10-LO}) are naively expected
to approximate well the couplings
at scales of the order of the relevant dynamics involved ($\mu \sim M_R$).
However they always carry an implicit error because of the uncertainty on $\mu$.
This theoretical uncertainty is rather important in couplings generated through scalar meson exchange,
such as $L_8^r(\mu)$~\cite{ChPTp4,L8-nlo}.
In the present case, it also has a moderate importance.
The size of the NLO corrections in $1/N_C$ to $L_{10}^r(\mu)$ and $C_{87}^r(\mu)$
can be estimated by regarding their variations with $\mu$.
These are respectively given by
\begin{equation}
\frac{\partial \, L_{10}^r}{\partial \ln \mu^2}
\,=\, -\, \frac{\Gamma_{10}}{32\pi^2}\,=\, 0.8\cdot 10^{-3} \, , \qquad
\frac{\partial \, C_{87}^r}{\partial \ln \mu^2}
\,=\, \frac{\Gamma_{87}^{(L)}}{32\pi^2}\,=\, -1.1\cdot 10^{-5} \, .
\end{equation}

At large $N_C$, a correlator that accepts an unsubtracted dispersion relation is determined by the position of the poles and the value of their residues, as shown in Eq.~(\ref{eq:LO}), which gives the general structure for $\Pi(t)$.
In our realization of the \rcht\ Lagrangian, this corresponds to a complete resonance saturation of the corresponding low-energy \chpt\ couplings. Operators of $\cO(p^{n>2})$ that only include Goldstone fields are absent in \rcht; they are generated (through resonance exchange) in the low-energy effective theory, \chpt, where the resonances have been integrated out. Thus, in Eqs.~\eqn{eq:L10-LO} we do not have any direct $\widetilde{L}_{10}$ or $\widetilde{C}_{87}$ contributions, where the tildes denote (non-existing) \rcht\ operators.

So far, we have been working within a $U(3)_L\otimes U(3)_R$ framework, but we are actually interested on the couplings of the standard $SU(3)_L\otimes SU(3)_R$ chiral theory. Thus, a matching between the two versions of \chpt\ must be performed~\cite{KL:00}. Nonetheless, on the contrary to what happens with other matrix elements
(e.g. the $S-P$ correlator~\cite{L8-nlo}), the spin--1 two-point functions do not gain contributions from the
$U(3)$--singlet chiral Goldstone; the $\eta_1$ does neither enter at tree-level nor in the one-loop correlators. Therefore, the corresponding LECs are identical in both theories at leading and next-to-leading order in $1/N_C$:  $L_{10}^r(\mu)^{U(3)} = L_{10}^r(\mu)^{SU(3)}$,
$C_{87}^r(\mu)^{U(3)}=C_{87}^r(\mu)^{SU(3)}$.

\subsection{$L_{10}^r(\mu)$ at NLO}

As a first determination of the chiral coupling $L_{10}$ at NLO, we give the expression obtained within the SRA approximation:
\begin{eqnarray}\label{eq:L10_SRA}
 L_{10}^r(\mu)|^{\mathrm{SRA}} &=&
 -\frac{F^2}{4}\! \left(\frac{1}{M_V^{r\, 2}}
 \!+\!\frac{1}{M_A^{r\,2}}\right) \! \left\{ 1+  \delta_{_{\rm NLO}}^{(1)} - \frac{M_V^{r\, 2}\,\delta_{_{\rm NLO}}^{(2)} }{M_V^{r\, 2}+M_A^{r\, 2}}  \right\}\! -\frac{1}{128\pi^2} \Bigg[ \ln \frac{M_V^2}{\mu^2} + \frac{1}{6}\nonumber \\
&& +\Frac{4 M_S^4}{M_V^4}-\Frac{7 M_S^2}{M_V^2}
+\left(-1 +\Frac{6 M_S^2}{M_V^2} -\Frac{9 M_S^4}{M_V^4}+\Frac{4 M_S^6}{M_V^6}  \right) \ln \left(1-\Frac{M_V^2}{M_S^2}\right)
\Bigg] ,
\end{eqnarray}
where we have used the relations in Eqs.~(\ref{dmr}) to eliminate the explicit dependence on the effective couplings $F_V^r$ and $G_V^r$, and the constraints of Eqs.~(\ref{masses}) to fix $M_A$ and $M_P$ at large $N_C$:
$M_A=M_V$ and $M_P=\sqrt{2}M_S$.

The needed input parameters are defined in the chiral limit.
We take the ranges~\cite{ChPTp4,polychromatic,PDG}
$M_V^r=M_V=(770\pm 5)$~MeV, 
$M_S=(1090\pm110)$~MeV and $F=(89\pm2)$~MeV.
Considering the importance of the axial-vector resonance field for the determination of this observable,
the mass provided by Ref.~\cite{PDG} is not satisfactory enough due to the large width of this meson.
We prefer to fix the value of $M_A$ in an indirect way,
by studying the decays of narrower resonances
like the $\rho(770)$. From the observed rates
$\Gamma (\rho^0 \rightarrow e^+ e^-)=(7.02\pm 0.13)$~keV~\cite{PDG}
and $\Gamma (\rho \rightarrow 2\pi) = (149.4 \pm 1.0)$~MeV~\cite{PDG}
one is able to estimate the values of the vector couplings,
$F_V=(155.7 \pm 1.5)$~MeV and $G_V=( 66.7 \pm 0.9)$~MeV.
These can both be used to recover the $a_1$ mass at large $N_C$:
one gets $M_A=(938\pm13)$~MeV from the WSR result in Eq.~(\ref{eq:FV_FA}), while
Eq.~(\ref{eq:GV}) gives the slightly larger range $M_A=(1160\pm 40)$~MeV.
Another large--$N_C$ determination of $M_A$ was obtained
in Ref.~\cite{jorge_vicent} from the study of the $\pi \rightarrow e \nu_e \gamma$ decay,
which yields
$M_A=(998\pm 49)$~MeV.  The renormalized mass $M_A^r$ can be recovered from
its large--$N_C$ value $M_A$ thanks to the experimental value of $F_V^r$ and
the WSR result in Eq.~(\ref{dmrbis}).
In spite of the dispersion of values for $M_A$,
the corresponding renormalized masses turn out to be always within the conservative range
$M_A^r=(1000\pm 50)$~MeV,  which we will take as our numerical input.

This gives for the SRA the numerical prediction
\begin{eqnarray}\label{eq:L10_SRA_num}
 L_{10}^r(\mu_0)|^{\mathrm{SRA}} &=&  (-5.2 \pm 0.4 ) \, \cdot \, 10^{-3}\, ,
\end{eqnarray}
being $\mu_0$ the usual renormalization scale, $\mu_0=770$~MeV.
Notice that in this expression we only consider the errors derived from
the experimental inputs. It does not include the systematic
uncertainties due to neglecting higher
resonance effects and the inconsistencies between form-factor constraints.

To asses the numerical impact of higher-mass resonances, we consider the results obtained
adding a second multiplet of vector and axial-vector states. The resulting value for the
chiral coupling  $L_{10}^r(\mu)$ takes the form
\begin{eqnarray}\label{eq:L10_2Res}
  L_{10}^r(\mu) &=&  -\frac{F^2}{4} \left(\frac{1}{M_V^{r\, 2}}+\frac{1}{M_A^{r\,2}}\right)  \left\{ 1+  \epsilon_{1}+\delta_{_{\rm NLO}}^{(1)} - \frac{M_V^{r\, 2}}{M_V^{r\, 2}+M_A^{r\, 2}}\left(\epsilon_{2} +\delta_{_{\rm NLO}}^{(2)} \right)  \right\} \nonumber \\
&& +\frac{F^2}{4M_{V'}^2} \left\{ \left(1+\frac{M_{V'}^2}{M_{A'}^2} \right) \epsilon_{1} -\frac{M_V^2}{M_{A'}^2} \epsilon_{2} \right\}   \nonumber \\
&& -\frac{1}{256\pi^2} \left\{ \frac{M_V^2}{M_{V'}^2-M_V^2} \left(1-\frac{M_V^2}{M_A^2} \right) \chi^{(1)}_{A\pi} +\frac{4M_V^2}{M_{V'}^2-M_V^2} \left(1-\frac{M_P^2}{M_S^2}\right) \chi^{(1)}_{P\pi} \right. \nonumber \\
&& \left.\qquad \quad +2\left(1-\frac{M_V^2}{M_A^2} \right) \chi_{V\pi}^{(1)} + 4\left(1-\frac{M_S^2}{M_P^2} \right) \chi_{S\pi}^{(1)}+2\ln \frac{M_V^2}{\mu^2} -\frac{16}{3} \right\} \, .
\end{eqnarray}
In the first line we have indicated the contribution coming from the tree-level exchange of the first multiplets $V$ and $A$, and Eqs.~(\ref{dmrbis}) have been used to fix the NLO couplings $F_V^r$ and $F_A^r$. The tree-level exchange of the second multiplets $V'$ and $A'$ generates the contributions shown in the second line, where $F_{V'}$ and $F_{A'}$ are expressed in terms of $\epsilon_{1}$ and $\epsilon_{2}$ [see Eqs.~(\ref{FVprime})]. The one-loop contribution, given in the third and fourth lines, is expressed in terms of known functions of the resonance masses, $\chi_{{m_1m_2}}^{(1)}$, which are given in appendix~B [see Eqs.~(\ref{LEApi1}), (\ref{LEPpi1}), (\ref{LEVpi1}) and (\ref{LESpi1})].

To obtain a numerical estimate, we need the masses of the lowest states,
the pion decay constant and the tiny corrections $\epsilon_{n}$,
which are related to $F_{V'}$ and $F_{A'}$. We take the following input
parameters~\cite{ChPTp4,polychromatic,PDG}
$M_V^r=M_V=(770\pm 5)$~MeV, $M_A^r=M_A=(1000\pm 50)$~MeV, $M_S=(1090\pm110)$~MeV,
$M_P=(1300\pm100)$~MeV and $F=(89\pm2)$~MeV.
The constraints
$\widetilde{\delta}_{_{\rm NLO}}^{(1)}=\widetilde{\delta}_{_{\rm NLO}}^{(2)}=0$
determine the $V'$ and $A'$ masses
[see Eqs.~(\ref{MVpMAp})].
Assuming the convergence of the Weinberg sum rules we consider
for the higher multiplet corrections
the range $\epsilon_i=0.0\pm 0.1$. At the usual $\chi PT$
renormalization scale $\mu_0=770$~MeV, one gets then
\begin{eqnarray}
\label{eq.L10}\label{eq:L10_2Res_num}
 L_{10}^r(\mu_0)&=& \left(-3.6 \pm 0.9 \pm 0.3\right)\cdot 10^{-3} \, ,
\end{eqnarray}
where the second error comes from the $\epsilon_i$ and the first one from the remaining inputs.

The assumed convergence of the Weinberg sum rules carries an implicit cancellation
between the tree-level contributions of the vector and axial-vector multiplets.
It is remarkable that this also leads to subtle cancelations between the $V\pi$ and $A\pi$
contributions.

\subsection{$C_{87}^r(\mu)$ at NLO}

The determination found within the SRA approximation reads
\begin{eqnarray}\label{eq:C87_SRA}
C_{87}^r(\mu)|^{\mathrm{SRA}}&=& \frac{F^4}{8}\! \left( \frac{1}{M_V^{r\,4}}
\!+\!  \frac{1}{M_V^{r\,2}M_A^{r\,2}}\! +\! \frac{1}{M_A^{r\,4}} \right)\!
\left\{ 1+  \delta_{_{\rm NLO}}^{(1)}  - \frac{M_V^{r\,2}\left(M_V^{r\,2}
+M_A^{r\,2}\right)\delta_{_{\rm NLO}}^{(2)}}{M_V^{r\,4}+M_A^{r\,4}+M_V^{r\,2}M_A^{r\,2}}  \right\}\nonumber \\
&&\mbox{}+\frac{F^2}{256\pi^2 M_V^2} \left[ \left(-2 +\Frac{9 M_S^2}{M_V^2} -\Frac{12 M_S^4}{M_V^4}
+\Frac{5 M_S^6}{M_V^6}  \right) \ln{\left(1-\Frac{M_V^2}{M_S^2}\right)} \nonumber \right.\\
&&\qquad \qquad \quad  \left.\mbox{} + \frac{1}{3} +2 \ln\frac{M_V^2}{\mu^2} +\Frac{5 M_S^4}{M_V^4}-\Frac{19 M_S^2}{2 M_V^2} -\Frac{M_V^2}{4M_S^2} \right] \, .
\end{eqnarray}
We have used again the relations in Eqs.~(\ref{dmr}) to fix $F_V^r$ and $F_A^r$ and the constraints coming from Eqs.~(\ref{masses}). Taking the same parameters than in the previous section for the SRA approximation, one finds, at $\mu_0=770$~MeV,
\begin{eqnarray}\label{eq:C87_SRA_num}
C_{87}^r(\mu_0)|^{\mathrm{SRA}}&=& (3.9\pm 0.6 )\, \cdot\, 10^{-5} \,.
\end{eqnarray}

Once the second multiplets $V'$ and $A'$ are included, the determination of the $\cO(p^6)$ chiral coupling takes the form:
\begin{eqnarray}\label{eq:C87_2Res}
 C_{87}^r(\mu)&=& \frac{F^4}{8} \left( \frac{1}{M_V^{r\,4}}\! +\!  \frac{1}{M_V^{r\,2}M_A^{r\,2}}\! + \!\frac{1}{M_A^{r\,4}} \right) \left\{ 1+\!  \delta_{_{\rm NLO}}^{(1)}\! +\!\epsilon_{1} \! -\! \frac{M_V^{r\,2}\left(M_V^{r\,2}\!+\!M_A^{r\,2}\right)}{M_V^{r\,4}\!+\!M_A^{r\,4}\!+\!M_V^{r\,2}M_A^{r\,2}} \right.\nonumber \\
&& \mbox{} \times  \left(\delta_{_{\rm NLO}}^{(2)}+\epsilon_{2} \right) \Bigg\}
-\!\frac{F^4}{8M_{V'}^2M_{A'}^2} \!\left\{
\left(1+\!\frac{M_{A'}^2}{M_{V'}^2}\!+\!\frac{M_{V'}^2}{M_{A'}^2} \right)
 \epsilon_{1} \!-\!\left(\frac{M_V^2}{M_{V'}^2}\! +\!\frac{M_V^2}{M_{A'}^2} \right)  \epsilon_{2} \right\} \nonumber \\
&&\!\!\!\!\!\!\!\!\!\!\!\!\mbox{} +\frac{F^2}{512\pi^2M_V^2} \left\{  \frac{M_V^2}{M_{V'}^2-M_V^2} \left(1-\frac{M_V^2}{M_A^2} \right) \chi_{A\pi}^{(2)} +\frac{4M_V^2}{M_{V'}^2-M_V^2}\left(1-\frac{M_P^2}{M_S^2} \right) \chi_{P\pi}^{(2)} \right. \nonumber \\
&&\left.\!\!\!\!\!\!\!\!\!\!\!\!\mbox{} +2\left(1-\frac{M_V^2}{M_A^2} \right) \frac{M_V^2}{M_A^2}\, \chi_{V\pi}^{(2)} +4\left(1-\frac{M_S^2}{M_P^2} \right) \frac{M_V^2}{M_{A'}^2}\, \chi_{S\pi}^{(2)} +4\ln \frac{M_V^2}{\mu^2} -\frac{26}{3}\right\}.
\end{eqnarray}
In the first and second lines we show the contributions coming from tree-level exchanges,
where again Eqs.~(\ref{dmrbis}) have been used to remove $F_V^r$ and $F_A^r$.
Again, $F_{V'}$ and $F_{A'}$ are expressed in terms of $\epsilon_{1}$ and
$\epsilon_{2}$. The third and fourth lines contain the one-loop contribution,
expressed through the known functions $\chi^{(2)}_{{m_1m_1}}$ which appear in appendix~B
[see Eqs.~(\ref{LEApi2}), (\ref{LEPpi2}), (\ref{LEVpi2}) and (\ref{LESpi2})].

Using the same parameters as for the $L_{10}^r(\mu)$ case, one gets the numerical estimate
\begin{eqnarray}\label{eq:C87_2Res_num}
C_{87}^r(\mu_0)&=& \left(2.2 \pm 1.0 \pm 0.4\right)\cdot 10^{-5}\,.
\end{eqnarray}
As in Eq.~\eqn{eq.L10}, the second error comes only from those in $\epsilon_1$
and $\epsilon_2$.

\section{Conclusions}\label{sec:summary}

The large--$N_C$ limit provides a very successful theoretical framework to understand the role of resonance
saturation in low-energy phenomenology \cite{polychromatic}. However, this limit is unable to pin down the
scale dependence of the \chpt\ couplings. Although this is a NLO effect in the $1/N_C$ expansion,
its numerical impact is very sizable.

In this paper we have presented a general method to determine the chiral couplings at NLO in $1/N_C$,
keeping full control of their renormalization-scale dependence. Through a one-loop calculation of appropriately chosen Green functions, within \rcht, one can get the needed NLO resonance contributions at low energies.
Using analyticity and unitarity, we avoid all technicalities associated with the renormalization procedure, reducing the calculation to much simpler dispersion relations. The QCD constraints at short distances provide a powerful tool to fix the corresponding subtraction constants.

\hspace*{0cm} From the theoretical analysis of the $\langle VV-AA\rangle$ correlator, we have obtained a
NLO prediction of the $O(p^4)$ coupling $L^r_{10}(\mu)$, which exactly
reproduces its right renormalization-scale dependence. Moreover, we have also determined the
$O(p^6)$ coupling $C^r_{87}(\mu)$ at the NLO, controlling its $\mu$ dependence up to small NNLO effects.

We have used the \rcht\ Lagrangian, within the SRA, to compute the one- and two-particle exchange contributions to
the absorptive part of the correlator. It is remarkable that, imposing a good short-distance
behaviour for the corresponding vector and axial-vector spectral functions, one fully determines
the relevant contributing form factors.
Using a dispersion relation,  we have reconstructed the
correlator, up to a term which has the same structure
as the tree-level one-particle contributions.
However, the stringent short-distance QCD constraints on $\Pi(t)$ have allowed
us to fix it in terms of resonance masses.
The low momentum expansion of the correlator $\Pi(t)$ reproduces the
right \chpt\ expression, with explicit values for the LECs $L^r_{10}(\mu)$ and $C^r_{87}(\mu)$ which only
depend on the resonance masses and the pion decay constant. The resulting analytical expressions for these
LECS are given in Eqs.~(\ref{eq:L10_SRA}) and (\ref{eq:C87_SRA}). Using the presently known information on the
resonance mass parameters, we obtain at $\mu= \mu_0=770$~MeV the numerical predictions in Eqs.~(\ref{eq:L10_SRA_num}) and (\ref{eq:C87_SRA_num}), respectively, where the main uncertainty originates in the input value of $M_A$.

To asses the impact of higher-mass states, we introduce a second multiplet
of vector and axial-vector states.
While this improves the theoretical description,
solving some conflicts between short-distance constraints
obtained from different form factors, it increases the number of parameters
making the numerical results more uncertain. Nevertheless, it is still possible
to obtain the explicit analytical predictions
in Eqs.~(\ref{eq:L10_2Res}) and (\ref{eq:C87_2Res}),
in terms of two small parameters $\epsilon_1$ and $\epsilon_2$, which are expected to be in the range
$\epsilon_i=0.0\pm 0.1$.
The corresponding numerical results at $\mu= \mu_0$, given
in Eqs.~(\ref{eq:L10_2Res_num}) and (\ref{eq:C87_2Res_num}),
have in both cases a smaller absolute value than the ones obtained
within the SRA.

%
%
\begin{figure}
\begin{center}
\begin{minipage}[c]{0.47\linewidth}\centering
\includegraphics[clip,scale=0.67]{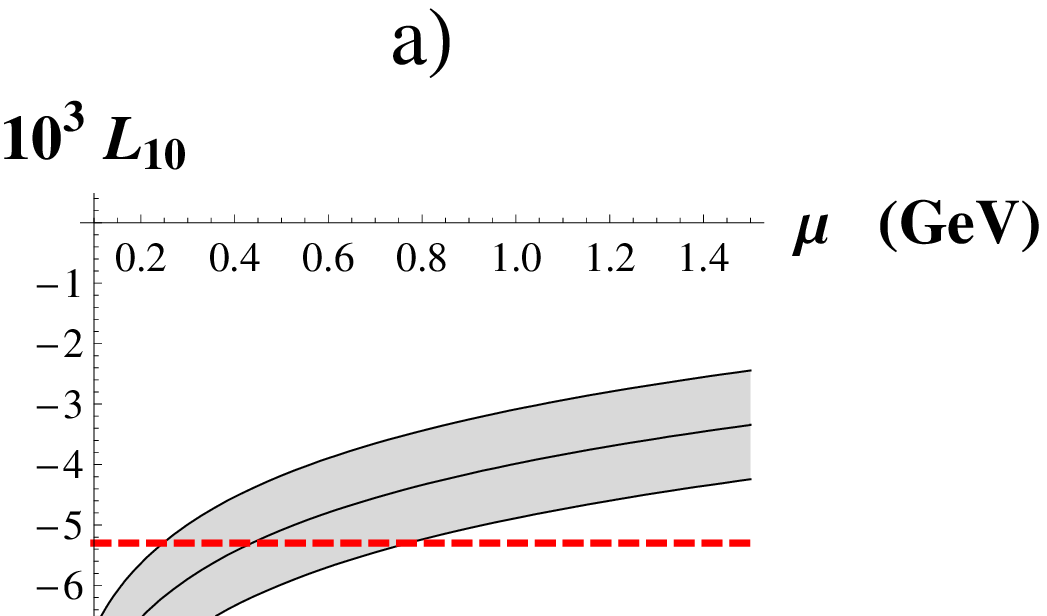}
\end{minipage}
\hfill
\begin{minipage}[c]{0.47\linewidth}\centering
\vskip .3cm
\includegraphics[clip,scale=0.67]{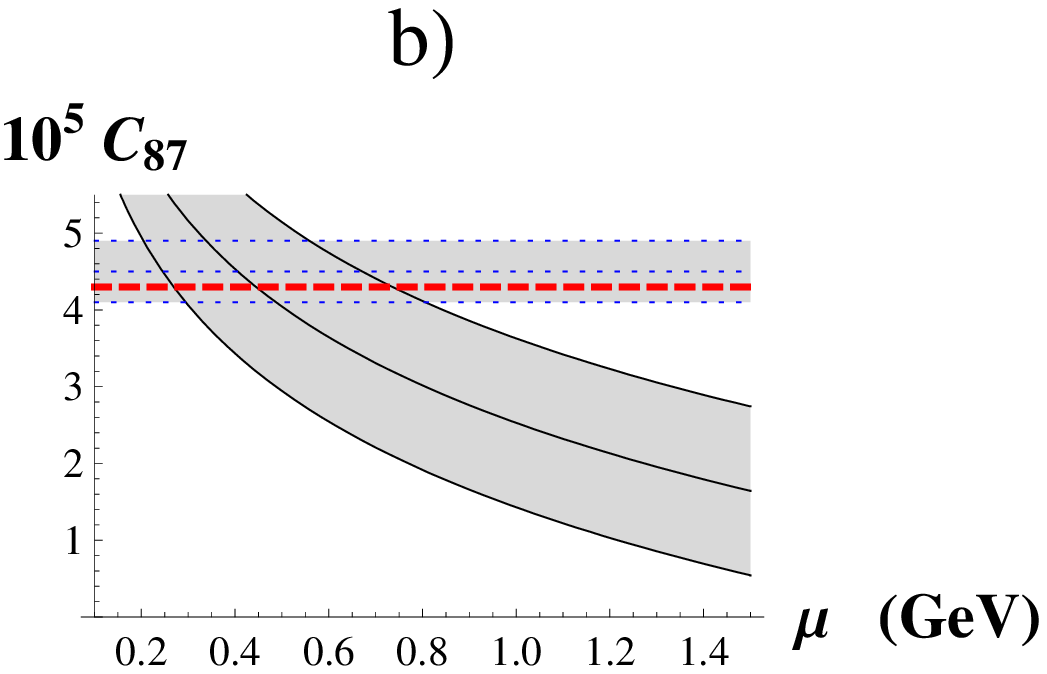}
\end{minipage}\vskip .2cm
\caption{\label{run-LECs}
a) Comparison of the NLO prediction for $L_{10}^r(\mu)$
as a function of $\mu$ (gray band) versus our large--$N_C$ estimate
(dashed);
b) NLO prediction for $C_{87}^r(\mu)$ (solid gray band)
compared to our LO estimate
(dashed) and the large--$N_C$ result from Ref.~\cite{C87-Peris} (dotted).
}
\end{center}
\end{figure}

We can combine the two numerical estimates of the LECs into our final results at $\mu= \mu_0$:
\begin{eqnarray}
 L_{10}^r(\mu_0)&=& \left(-4.4 \pm 0.9\right)\cdot 10^{-3} \, ,
\nn\\
C_{87}^r(\mu_0)&=& \left(3.1 \pm 1.1\right)\cdot 10^{-5}\, .
\end{eqnarray}
The central values lie in between the two determinations, but we have kept the larger error bars of
Eqs.~(\ref{eq:L10_2Res_num}) and (\ref{eq:C87_2Res_num}) (adding in quadrature the two uncertainties).
Figure~\ref{run-LECs} shows the corresponding predictions as functions of the
renormalization scale $\mu$. Also shown are
the large--$N_C$ results and the recent Pade estimate of
Ref.~\cite{C87-Peris}, which cannot incorporate
the dependence on the scale $\mu$. The figure shows that
the leading-order approximations agree with our
NLO results for values of the renormalization scale around
$\mu\sim 0.5$~GeV.

The ideas discussed in this article can be applied to generic Green functions, which opens
a way to investigate other chiral LECs at NLO in the large--$N_C$ expansion.
In particular, it looks feasible to analyze the couplings $L_5$ and $L_9$
with similar techniques. Further work in this direction
is in progress.

\section*{Acknowledgments}

This work has been supported by the
Spanish Ministry of Education, under grants FPA2007-60323 and CSD2007-00042 (Consolider Project CPAN),
and by the EU MRTN-CT-2006-035482 (FLAVIAnet).

\newpage

\appendix

\section{Two-meson form factors and their short-distance constraints}\label{ap:A}

\newcounter{catilina}
\renewcommand{\thetable}{\Alph{catilina}.\arabic{table}}
\setcounter{catilina}{1}
\setcounter{table}{0}

In this appendix all two-body form factors that can be found in the even-intrinsic-parity sector of the R$\chi$T in the SRA are analysed, following the ideas of section~3. Furthermore the needed form factors for the $V-A$ correlator with higher multiplets $V'$ and $A'$ are studied.

The following items are presented for each form factor:
\begin{enumerate}
\item The form factor(s) is (are) defined through the corresponding matrix element.
\item The expression of the form factor(s) is (are) shown.
\item Using the optical theorem, the spectral function is given
 in terms of the form factors.
\item The constraints found by imposing a good high-energy behaviour of the spectral function.
\item  Once the constraints are imposed, the
well behaved form factor(s) is (are) presented again and quoted with a tilde.
\end{enumerate}
Notice that when $R^0_{I=0}$  or $\eta$
is written, we refer to the singlet in the $U(2)$ case.
The following usual notation is employed throughout the section :
\begin{equation}
\lambda\left(a,b,c\right) = a^2+b^2+c^2-2ab-2ac-2bc  \, , \qquad \sigma_M =
\sqrt{1-4M^2/t}
\, .\nonumber
\end{equation}

\begin{center}
\begin{longtable}{|>{$}c<{$}|>{$}c<{$}|}
%
%
\caption{Vector form factors: amplitudes, definitions, spectral functions and constraints.} \label{tableVFF} \\
\hline
\mathbf{1} & \mathcal{V}^\mu_1=\bra \pi^0 (p_1) \pi^- (p_2) | \bar{d}\gamma^\mu u | 0\ket = \sqrt{2}\, \mathcal{F}_{\pi\pi}^{\,v} (t)\, (p_2-p_1)^\mu \phantom{\bigg(} \\
\hline
\hline
\mathcal{F}_{\pi\pi}^{\,v}& \phantom{\Bigg(}1 + \Frac{F_VG_V}{F^2} \Frac{t}{M_V^2-t} \\
\hline
\mathrm{Im}  \Pi_{\pi\pi}  & \Frac{\theta(t)}{24\pi} |\mathcal{F}_{\pi\pi}^{\,v} (t)|^2 \phantom{\Bigg(}\\
\hline
\multicolumn{2}{|>{$}c<{$}|}{
 F_VG_V=F^2 \qquad \qquad \longrightarrow \qquad \qquad
\mathcal{\widetilde{F}}_{\pi\pi}^{\,v} = \Frac{M_V^2}{M_V^2-t}\phantom{\Bigg(}} \\
\hline
\multicolumn{2}{c}{} \\
%
%
\hline
\mbf{2} & \mathcal{V}^\mu_2 = \bra A^0_{I=1} (p_A,\varepsilon) \pi^- (p_\pi) | \bar{d}\gamma^\mu u | 0\ket \phantom{\bigg(}\\
\hline
\hline
\mathcal{V}^\mu_2 & \Frac{i\, \sqrt{2}}{M_A} \Big\{
( q\varepsilon^* \, p_A^\mu \, -\, qp_A\, {\varepsilon^*}^\mu )\, \mathcal{F}^{\,v}_{A\pi} (t)+( q\varepsilon^* \, p_\pi^\mu \, -\, qp_\pi {\varepsilon^*}^\mu )\, \mathcal{G}^{\,v}_{A\pi} (t) \Big\} \phantom{\Bigg(} \\
\hline
\end{longtable}
\end{center}

\begin{center}
\begin{longtable}{|>{$}c<{$}|>{$}c<{$}|}
%
%
\caption{Axial form factors: amplitudes, definitions, spectral functions and constraints.}\label{tableAFF}\\
\hline
\mathbf{1} & \mathcal{A}^\mu_1= \bra V^0_{I=1} (p_V,\varepsilon) \pi^- (p_\pi) | \bar{d}\gamma^\mu \gamma_5 u | 0\ket \phantom{\bigg(}\\ \hline
\mathcal{A}^\mu_1 \phantom{\Bigg(}& \Frac{i\, \sqrt{2}}{M_V} \Big\{( q\varepsilon^* \, p_V^\mu \, -\, qp_V\, {\varepsilon^*}^\mu )\, \mathcal{F}^{\,a}_{V\pi} (t) +( q\varepsilon^* \, p_\pi^\mu \, -\, qp_\pi {\varepsilon^*}^\mu )\, \mathcal{G}^{\,a}_{V\pi} (t) \Big\} \\ \hline
\mathcal{F}^{\,a}_{V\pi} \phantom{\Bigg(} & -\Frac{F_V}{F}+\Frac{2G_V}{F}-\Frac{2G_V}{F}\Frac{M_V^2}{t} + \Frac{F_A}{F}\Frac{t}{M_A^2-t}\Big[(-\Frac{2M_V^2}{t}+2) \lambda^{VA}_2  \\ \phantom{\bigg(}
&+(\Frac{M_V^2}{t}-1) \lambda^{VA}_4+ (\Frac{2M_V^2}{t}-2) \lambda^{VA}_5 \Big]  \\ %
\mathcal{G}^{\,a}_{V\pi}\phantom{\Bigg(}  & -\Frac{2G_V}{F}\Frac{M_V^2}{q^2}+ \Frac{2F_A}{F}\Frac{M_V^2}{M_A^2-q^2} \Big[ -2\lambda^{VA}_2+\lambda^{VA}_3 \Big] \\ \hline
\mathrm{Im}   \Pi_{V\pi} & \phantom{\bigg(}\mathrm{Im} \Pi_{A\pi}\left( M_A \rightarrow M_V, \mathcal{F}^v_{A\pi} \rightarrow \mathcal{F}^a_{A\pi}, \mathcal{G}^v_{A\pi} \rightarrow \mathcal{G}^a_{V\pi} \right) \\ \hline
\multicolumn{2}{|>{$}c<{$}|}{\phantom{\Bigg(}
2\lambda^{VA}_2-\lambda^{VA}_4-2\lambda^{VA}_5=-\Frac{F_V}{F_A}+\Frac{2G_V}{F_A}  \qquad
-2\lambda^{VA}_2+\lambda^{VA}_3=-\Frac{G_V}{F_A}  \label{AVG}  } \\
\hline
\end{longtable}
\end{center}

\begin{center}
\begin{longtable}{|>{$}c<{$}|>{$}c<{$}|}
%
%
\caption{Scalar form factors: amplitudes, definitions, spectral functions and constraints.}\label{tableSFF}\\
\hline
\mathbf{1} & \mathcal{S}^\mu_1= \bra \eta (p_\eta) \pi^- (p_\pi) | \bar{d} u | 0\ket  =  \mathcal{F}^{\,s}_{\pi \eta} (t) \phantom{\bigg(} \\ \hline
\mathcal{F}^{\,s}_{\pi \eta} \phantom{\Bigg(} &\sqrt{2} \, B_0 \left( 1+4\,\Frac{c_m c_d}{F^2}\, \Frac{t}{M_S^2-t} \right) \\ \hline
\mathrm{Im} \Pi_{\pi\eta} & \phantom{\Bigg(} \theta(t) \Frac{1}{16\pi} |\mathcal{F}^{\,s}_{\pi \eta} (t)|^2 \\ \hline
\multicolumn{2}{|>{$}c<{$}|}{\phantom{\Bigg(}
4\,c_d\,c_m=F^2 \qquad  \longrightarrow  \qquad
\mathcal{\widetilde{F}}^{\,s}_{\pi \eta} (t) =\sqrt{2}B_0\Frac{M_S^2}{M_S^2-t}} \\ \hline
\multicolumn{2}{c}{} \\ \hline
%
%
\mbf{2} & \mathcal{S}^\mu_2= \bra A^0_{I=0} (p_A,\varepsilon) \pi^- (p_S)  | \bar{d} u | 0\ket =
\Frac{i}{M_A}  \, q\varepsilon^*  \,\mathcal{F}^{\,s}_{A\pi} (t) \phantom{\Bigg(}\\ \hline
\mathcal{F}^{\,s}_{A\pi}\phantom{\Bigg(} & -\Frac{8\, B_0 \,c_m \, \lambda^{SA}_1}{F} \,\Frac{M_A^2}{M_S^2-t}  \\
\hline
\end{longtable}
\end{center}

\begin{center}
\begin{longtable}{|>{$}c<{$}|>{$}c<{$}|}
%
%
\caption{Pseudoscalar form factors: amplitudes, definitions, spectral functions and constraints.}\label{tablePFF}\\
\hline
\mathbf{1}\phantom{\Bigg(} & \mathcal{P}^\mu_1=
\bra \pi^0 (p_\pi) V^- (p_V,\varepsilon) |i \bar{d}\gamma_5 u | 0\ket =
\Frac{1}{M_V}  \, q\varepsilon^*  \,\mathcal{F}^{\,p}_{V\pi} (t) \\ \hline
\mathcal{F}^{\,p}_{V\pi}\phantom{\Bigg(} & -\Frac{2\, B_0}{F} \, \left( \sqrt{2}G_V \Frac{M_V^2}{t} +4d_m\, \lambda^{PV}_1\Frac{M_V^2}{M_P^2-t}    \right)  \\ \hline
\mathrm{Im}  \Pi_{V\pi} &\phantom{\Bigg(} \theta(t-M_V^2) \,\Frac{\left( t-M_V^2 \right)^3 }{64\pi M_V^4 t}\, |\mathcal{F}^{\,p}_{V\pi}|^2 \\ \hline
\multicolumn{2}{|>{$}c<{$}|}{\phantom{\Bigg(}
-\sqrt{2} G_V + 4d_m \, \lambda^{PV}_1 = 0  \qquad \longrightarrow \qquad
\mathcal{\widetilde{F}}^{\,p}_{V\pi} (t)=-\Frac{2\sqrt{2}B_0 \,G_V}{F} \Frac{M_V^2M_P^2}{(M_P^2-t)t} } \\ \hline
\multicolumn{2}{c}{} \\ \hline
%
%
\mbf{2}\phantom{\bigg(} & \mathcal{P}^\mu_{2} =
\bra S^0_{I=0} (p_S) \pi^- (p_\pi) | i \bar{d} \gamma_5 u | 0\ket =\mathcal{F}^{\,p}_{S\pi} (t)  \\ \hline
\mathcal{F}^{\,p}_{S\pi} \phantom{\Bigg(}& \Frac{4B_0\,c_m}{F} -\Frac{2B_0\,c_d}{F} \Frac{t-M_S^2}{t} \,+\,\Frac{4B_0\,d_m}{F}\Frac{M_S^2-t}{M_P^2-t} \lambda^{SP}_1 \\
\hline
\end{longtable}
\end{center}

\begin{center}
\begin{longtable}{|>{$}c<{$}|>{$}c<{$}|}
\caption{Form Factors with extra multiplets $V'$ and $A'$.}\label{tableextra}\\
\hline
%
%
\multicolumn{2}{|c|}{\phantom{\bigg(} {\bf Vector form factor to $\pi\pi$}} \\ \hline
\mathcal{F}_{\pi\pi}^{\,v} &\phantom{\Bigg(}1\,+\,\Frac{F_VG_V}{F^2} \Frac{t}{M_V^2-t} +\,\Frac{F_{V'}G_{V'}}{F^2} \Frac{t}{M_{V'}^2-t} \\ \hline
\multicolumn{2}{|>{$}c<{$}|}{\phantom{\bigg(}
F_V\, G_V + F_{V'}\, G_{V'} \,=\, F^2 }   \\ \hline
\mathcal{\widetilde{F}}_{\pi\pi}^{\,v} & \phantom{\Bigg(} \Frac{M_V^2}{M_V^2-t}+\cO(\epsilon) \\ \hline
\multicolumn{2}{c}{} \\ \hline
%
%
\multicolumn{2}{|c|}{\phantom{\bigg(}{\bf Vector form factor\,\footnote{
We use here the constraint coming from the pseudoscalar form factor to $V\pi$ (Table \ref{tablePFF}).} to $P\pi$}
} \\ \hline
\mathcal{F}^{\,v}_{P\pi} & \phantom{\Bigg(}\Frac{2 \lambda^{PV}_1 F_V}{F} \Frac{1}{M_V^2-t}+\Frac{2 \lambda^{PV'}_1 F_{V'}}{F} \Frac{1}{M_{V'}^2-t} \\
\hline
\end{longtable}
\end{center}

\section{One-loop corrections $\Pi (t)|_{m_1,m_2}$}\label{ap:B}

\subsection{Single resonance approximation}

In this appendix we show the explicit form of
the one-loop corrections generated by the considered two-particle
absorptive cuts,
which have been calculated by using the dispersive method discussed in section~4.
Note that the different  $\Pi(t)|_{R\pi}$ include the factor 2 that accounts
that, for instance, the axial correlator gains contributions
both from the $\rho^0 \pi^-$ and $\rho^- \pi^0$ channels.
These expressions can be simplified by means of the large--$N_C$ relations
in Eqs.~(\ref{eq:FV_FA}),~(\ref{eq:GV}) and (\ref{eq:cd}),
which relate the resonance masses and the couplings  $F_V$, $G_V$, $F_A$, $c_d$:
\begin{eqnarray}
 \Pi|_{V\pi}\, &=&\,\,
\,  \Frac{n_f}{2}\,  \Frac{1}{24\, \pi^2}\,
\left(1-\Frac{M_V^2}{M_A^2}\right)  \,
\, \left(\Frac{M_A^2}{M_A^2\, -\, t}\right)^2
\nn \\
&& \times \left\{
-\Frac{2M_V^4}{t^2} - \Frac{8 M_V^2}{t}
+\left(\Frac{M_A^2}{M_V^2} -6 +\Frac{25 M_V^2}{M_A^2} +\Frac{8M_V^4}{M_A^4}\right)
\right.
\nn\\
&&\qquad \qquad \qquad
+\Frac{t}{M_V^2} \left(-1 +\Frac{6M_V^2}{M_A^2} -\Frac{17 M_V^4}{M_A^4}
-\Frac{6M_V^6}{M_A^6}\right)
\nn\\
&& \quad
+
  \left[
  \left(7-\Frac{30 M_V^2}{M_A^2} +\Frac{21 M_V^4}{M_A^4} +\Frac{8M_V^6}{M_A^6}\right)
  \right.
\nn\\
&&\qquad \qquad
  \left.
  +\Frac{t}{M_V^2}\left( -1+\Frac{15M_V^4}{M_A^4} -\Frac{14 M_V^6}{M_A^6} -\Frac{6 M_V^8}{M_A^8}\right)
  \right] \, \ln{\left(\Frac{M_A^2}{M_V^2}-1\right)}
\nn\\
&&\quad
+
\left.
  \left[
  -\Frac{2M_V^6}{t^3}-\Frac{7M_V^4}{t^2}+\Frac{15 M_V^2}{t} -7 +\Frac{t}{M_V^2}
  \right]\, \ln{\left(1-\Frac{t}{M_V^2}\right)}
\right\}
\,  , \label{totc4b}\\
 \Pi|_{A\pi}\, &=&\,
\Frac{n_f}{2}\, \Frac{1}{24\pi^2} \,\left(\Frac{M_A^2}{M_V^2}-1\right)
\, \left( \Frac{M_V^2}{M_V^2-t}\right)^2
\nn \\
&&
\times
 \left\{
 \Frac{M_A^2}{t} + \left( -\Frac{M_V^2}{M_A^2} -4-\Frac{3M_A^2}{M_V^2}\right)
 + \Frac{t}{M_A^2} \left(1 +\Frac{4M_A^2}{M_V^2} +\Frac{2 M_A^4}{M_V^4}\right)
 \right.
\nn \\
&&
 +
   \left[
   \left(3-\Frac{6M_A^2}{M_V^2} - \Frac{3 M_A^4}{M_V^4}\right)
   +\Frac{t}{M_A^2}\left(1+\Frac{3M_A^4}{M_V^4} +\Frac{2 M_A^6}{M_V^6}\right)
   \right]   \ln{\left(1-\Frac{M_V^2}{M_A^2}\right)}
\nn \\
&& \qquad
\left.
 - \left(1-\Frac{M_A^2}{t}\right) \, \left(\Frac{M_A^2}{t}+4+\Frac{t}{M_A^2}\right)
   \ln{\left( 1-\Frac{t}{M_A^2}\right)}
 \right\}
\, \label{totc2b}
\\
\Pi|_{S\pi}\, &=&\,\,
\,\Frac{n_f}{2}\Frac{1}{12\, \pi^2}
\, \left(  1-\Frac{M_S^2}{M_P^2} \right)\,
\left(\Frac{M_A^2}{M_A^2\, -\, t}\right)^2 \,
\nn \\
&& \times \left\{
- \Frac{M_S^4}{t^2} + \Frac{5 M_S^2}{2 t}
+ \left( 1-\Frac{7 M_S^2}{M_A^2} + \Frac{4 M_S^4}{M_A^4}\right)
+ \Frac{t M_S^2}{2 M_A^4} \left(9-\Frac{6 M_S^2}{M_A^2}
                                   -\Frac{2M_A^2}{M_S^2}\right)
\right. \nn
\\
&&\qquad
- \left(1-\Frac{M_S^2}{M_A^2}\right)^2
  \left[1-\Frac{4M_S^2}{M_A^2}+\Frac{3 t M_S^2}{M_A^4}\right]
  \ln{\left(\Frac{M_A^2}{M_S^2}-1\right)}
\nn \\
&& \qquad \left.
+
  \left( 1-\Frac{M_S^2}{t} \right)^3 \ln{\left(1-\Frac{t}{M_S^2}\right)}
\right\}
\, ,\label{totc5b}
\\
 \Pi(t)|_{P\pi}\, &=&\,\,\, 0\, . 
\label{totc3b}
\end{eqnarray}

\subsection{Contributions from extra multiplets $V'$, $A'$}

Now we give the explicit expressions of the one-loop resonance corrections in the case of considering an extra multiplet for $V'$ and $A'$. In the case of the $V\pi$ cut, the result is the same as in the SRA, {\it i.e.} Eq.~(\ref{totc4b}). For the case of the $S\pi$ one can use the result in the SRA, with only changing $M_A$ to $M_{A'}$ everywhere. For the other ones, one finds
\begin{eqnarray}
\Pi |_{P\pi} &=& \frac{n_f}{2} \frac{1}{12\pi^2} \left(\frac{M_P^2}{M_S^2}-1\right)
\left( \frac{M_V^2M_{V'}^2}{\left(M_V^2-t\right)\left(M_{V'}^2-t\right)} \right)^2 \frac{M_V^2}{M_{V'}^2-M_V^2} \nonumber \\
&& \times \left\{ \left( 1-\frac{M_V^2}{M_{V'}^2}\right)\left( 1 - \frac{t}{M_{V'}^2} \right) \left( 1-\frac{t}{M_V^2}\right) \left[ -\frac{M_P^4}{M_V^2M_{V'}^2} - \frac{M_P^4 M_{V'}^2}{M_V^6}+\frac{2 M_P^2}{M_V^2}\right. \nonumber \right. \\
&& \left.  +\frac{2M_P^2M_{V'}^2}{M_V^4} -\frac{2M_{V'}^2}{M_V^2} +\frac{t}{M_V^2} \left( 1+\frac{M_{V'}^2}{M_V^2}-\frac{4M_P^2}{M_V^2}+\frac{M_P^4}{M_V^4} +\frac{M_P^4}{M_V^2M_{V'}^2} \right) \right] \nonumber \\
&& - \left(1\!-\!\frac{t}{M_{V'}^2}\right)^2 \!\left(1-\frac{M_P^2}{M_V^2}\right)^2 \!\left[ 1+\frac{M_{V'}^2}{M_V^2}-\frac{4M_P^2}{M_V^2}+\frac{2M_P^2M_{V'}^2}{M_V^4} +\frac{t}{M_V^2} \!\left(-\frac{2M_{V'}^2}{M_V^2} \right. \right. \nonumber \\
&& \left. \left. +\frac{3M_P^2}{M_V^2}-\frac{M_{V'}^2 M_P^2}{M_V^4}\right)\right] \ln \left( 1-\frac{M_V^2}{M_P^2} \right)+\left(1-\frac{t}{M_V^2}\right)^2\left(1-\frac{M_P^2}{M_{V'}^2}\right)^2 \nonumber \\
&&\left[ 1+\frac{2M_P^2}{M_{V'}^2}-\frac{4M_P^2}{M_V^2}+\frac{M_{V'}^2}{M_V^2}+\frac{t}{M_{V'}^2}\left(-2-\frac{M_P^2}{M_{V'}^2}+\frac{3M_P^2}{M_V^2} \right) \right] \ln \left( \frac{M_{V'}^2}{M_P^2}-1\right) \nonumber \\ &&\left.
-\left( 1-\frac{M_V^2}{M_{V'}^2}\right)^3\left(\frac{t-M_P^2}{M_V^2}\right)^3 \frac{M_{V'}^2}{t} \ln \left( 1-\frac{t}{M_P^2} \right) \right\} \, , \\
&& \nonumber \\
 \Pi|_{A\pi}&=& \frac{n_f}{2} \frac{1}{24\pi^2} \left(1-\frac{M_V^2}{M_A^2}\right)
\left( \frac{M_V^2M_{V'}^2}{\left(M_V^2-t\right)\left(M_{V'}^2-t\right)} \right)^2 \frac{M_V^2}{M_{V'}^2-M_V^2} \nonumber \\
&&\times \left\{ \!\left(1-\frac{M_V^2}{M_{V'}^2}\right)\! \left(1-\frac{t}{M_V^2}\right)\! \left(1-\frac{t}{M_{V'}^2} \right)\! \left[ \frac{M_A^2}{t} \left( \frac{M_A^6}{M_V^4M_{V'}^2} +\frac{M_A^6M_{V'}^2}{M_V^8}-\frac{2M_A^6}{M_V^6} \right. \right. \right. \nonumber \\
&& \left. +\frac{2M_A^4}{M_V^4}-\frac{2M_{V'}^2M_A^4}{M_V^6}+ \frac{M_{V'}^2M_A^2}{M_V^4} \!\right)\! -\frac{M_A^8}{M_V^8}\!\left(\!1+\frac{M_V^2}{M_{V'}^2}\!\right) \left(\! -3+\frac{2M_V^2}{M_{V'}^2}+ \frac{2M_{V'}^2}{M_V^2}\!\right) \nonumber \\
&& +\frac{2M_A^6}{M_V^4M_{V'}^2}+\frac{8M_A^6M_{V'}^2}{M_V^8} -\frac{7M_A^4}{M_V^4} - \frac{10M_A^4M_{V'}^2}{M_V^6} +\frac{2M_A^2 M_{V'}^2}{M_V^4} -\frac{M_{V'}^4}{M_V^4} \nonumber \\
&&+\frac{t}{M_V^2} \left( \frac{2M_A^8}{M_V^4M_{V'}^4}-\frac{2M_A^8}{M_V^6M_{V'}^2}+ \frac{2M_A^8}{M_V^8} -\frac{2M_A^6}{M_V^4M_{V'}^2}-\frac{8M_A^6}{M_V^6}+\frac{16M_A^4}{M_V^4} +\frac{2M_A^2}{M_V^2}\right. \nonumber \\
&& \left. \left. -\frac{2M_A^2M_{V'}^2}{M_V^4}+\frac{M_{V'}^4}{M_V^4}\! \right)\right]\! + \!\left(\!1-\frac{t}{M_V^2}\!\right)^2\!\left(\!1-\frac{M_A^2}{M_{V'}^2}\!\right)^3 \left[ \!\frac{5M_A^4}{M_V^4} -\frac{3M_A^4}{M_V^2M_{V'}^2} -\frac{M_A^2}{M_V^2} \right. \nonumber \\
&& \left. -\frac{5M_A^2M_{V'}^2}{M_V^4} + \frac{t}{M_V^2} \left(-1-\frac{4M_A^4}{M_V^2M_{V'}^2} +\frac{2M_A^4}{M_{V'}^4} +\frac{3M_A^2}{M_{V'}^2} +\frac{3M_A^2}{M_V^2} +\frac{M_{V'}^2}{M_V^2} \right) \right] \nonumber \\
&& \times \ln \left( \frac{M_{V'}^2}{M_A^2}-1 \right) + \left(1-\frac{M_V^2}{M_{V'}^2} \right) \left( 1-\frac{M_A^2}{t} \right) \left[ -\frac{M_A^2}{t} \left( \frac{M_A^6}{M_V^4M_{V'}^2}+\frac{M_A^6M_{V'}^2}{M_V^8} \right. \right. \nonumber \\
&& \left. -\frac{2M_A^6}{M_V^6} +\frac{2 M_A^4}{M_V^4} -\frac{2M_A^4M_{V'}^2}{M_V^6} +\frac{M_A^2 M_{V'}^2}{M_V^4}\! \right)\! +\!\left(\!1-\frac{M_V^2}{M_{V'}^2}\!\right) \!\left(\!-\frac{6M_A^6}{M_V^6} +\frac{4M_A^6M_{V'}^2}{M_V^8}\! \right) \nonumber \\
&& +\frac{4M_A^4}{M_V^4} -\frac{2M_A^4M_{V'}^2}{M_V^6} -\frac{4M_A^2 M_{V'}^2}{M_V^4} +\frac{t}{M_V^2} \left(-\frac{9M_A^4}{M_V^2M_{V'}^2}+\frac{14M_A^4}{M_V^4} -\frac{6M_A^4 M_{V'}^2}{M_V^6} \right. \nonumber \\
&& \left. +\frac{10M_A^2}{M_V^2} -\frac{2M_A^2M_{V'}^2}{M_V^4} -\frac{M_{V'}^2}{M_V^2} \!\right)\! +\frac{t^2}{M_V^4} \!\left(\!- \frac{2M_A^2}{M_{V'}^2}-\frac{6M_A^2}{M_V^2}+\frac{4M_A^2M_{V'}^2}{M_V^4} +\frac{2M_{V'}^2}{M_V^2} \!\right) \nonumber \\
&& \left. -\frac{t^3 M_{V'}^2}{M_V^8} \right] \ln \left( 1-\frac{t}{M_A^2} \right) +\frac{M_A^2}{M_V^2} \left( 1 -\frac{t}{M_{V'}^2} \right)^2 \left[ \frac{5M_A^8}{M_V^8} -\frac{3M_A^8M_{V'}^2}{M_V^{10}} -\frac{28M_A^6}{M_V^6} \right. \nonumber \\
&& +\frac{16M_A^6M_{V'}^2}{M_V^8} +\frac{45M_A^4}{M_V^4} -\frac{21M_A^4M_{V'}^2}{M_V^6} -\frac{14M_A^2}{M_V^2} -\frac{6M_A^2M_{V'}^2}{M_V^4} -2 + \frac{8M_{V'}^2}{M_V^2}  \nonumber \\
&& +\frac{t}{M_V^2} \left( -\frac{4M_A^8}{M_V^8} +\frac{2M_A^8M_{V'}^2}{M_V^{10}} +\frac{21 M_A^6}{M_V^6} -\frac{9M_A^6M_{V'}^2}{M_V^8} -\frac{30M_A^4}{M_V^4} +\frac{6M_A^4 M_{V'}^2}{M_V^6}  \right. \nonumber \\
&& \left. \left.\left.+\frac{7M_A^2}{M_V^2}  +\frac{13 M_A^2 M_{V'}^2}{M_V^4} -\frac{6M_{V'}^2}{M_V^2}\right)  \right] \ln \left( 1-\frac{M_V^2}{M_A^2} \right) \right\} \, .
\end{eqnarray}

In section~4.2 the high-energy behaviour of the correlator is expressed in terms of the functions $\xi^{(m)}_{m_1m_2}$, which read

\begin{eqnarray}
 \xi_{A\pi}^{(1)} &=& \left(1-\frac{M_V^2}{M_{V'}^2}\right) \left( \frac{2M_A^8}{M_V^6M_{V'}^2} -\frac{2M_A^8}{M_V^8} -\frac{4M_A^2M_{V'}^4}{M_V^6} +\frac{2M_A^8M_{V'}^2}{M_V^{10}} -\frac{8M_A^6M_{V'}^2}{M_V^8}  -\frac{2M_A^6}{M_V^6} \right. \nonumber \\
&& \left. +\frac{16M_A^4M_{V'}^2}{M_V^6} +\frac{2M_A^2M_{V'}^2}{M_V^4} +\frac{M_{V'}^6}{M_V^6} \right) +\left(1-\frac{M_{V'}^2}{M_V^2} \right) \frac{M_{V'}^4}{M_V^4} \ln \frac{M_V^2}{M_A^2} \nonumber \\
&& +\left(1-\frac{M_A^2}{M_{V'}^2} \right)^3 \left( \frac{2M_A^4}{M_V^4} -\frac{4M_A^4M_{V'}^2}{M_V^6} +\frac{3M_A^2M_{V'}^2}{M_V^4} +\frac{3M_A^2 M_{V'}^4}{M_V^6} +\frac{M_{V'}^6}{M_V^6} -\frac{M_{V'}^4}{M_V^4} \right)\nonumber \\
&& \times  \ln \left(\frac{M_{V'}^2}{M_A^2} -1\right) + \left( -\frac{4M_A^{10}}{M_V^{10}}+ \frac{2M_A^{10}M_{V'}^2}{M_V^{12}} -\frac{9M_A^8M_{V'}^2}{M_V^{10}} +\frac{21M_A^8}{M_V^8} -\frac{30M_A^6}{M_V^6} \right. \nonumber \\
&& \left. +\frac{6M_A^6M_{V'}^2}{M_V^8} +\frac{7M_A^4}{M_V^4} +\frac{13M_A^4 M_{V'}^2}{M_V^6} -\frac{6M_A^2M_{V'}^2}{M_V^4} \right) \ln \left( 1-\frac{M_V^2}{M_A^2} \right) \, , \\
\nonumber \\
\xi_{A\pi}^{(2)} &=& \left(1\!-\!\frac{M_V^2}{M_{V'}^2} \right) \left( \frac{M_A^8}{M_V^8} \!+\!\frac{M_A^8M_{V'}^2}{M_V^{10}} \!-\!\frac{10M_A^6M_{V'}^2}{M_V^8} \!+\!\frac{9M_A^4M_{V'}^2}{M_V^6} \!+\!\frac{6M_A^4M_{V'}^4}{M_V^8}\! +\!\frac{2M_A^2 M_{V'}^2}{M_V^4}\right. \nonumber \\
&& \left. -\frac{3M_A^2M_{V'}^6}{M_V^8} +\frac{M_{V'}^8}{M_V^8} \right) +\left(1-\frac{M_A^2}{M_{V'}^2}\right)^3 \frac{M_{V'}^2}{M_V^2} \left( \frac{M_A^4}{M_V^4} -\frac{3M_A^4M_{V'}^2}{M_V^6}+\frac{5M_A^2M_{V'}^2}{M_V^4} \right. \nonumber \\
&& \left. +\frac{M_A^2M_{V'}^4}{M_V^6} +\frac{2M_{V'}^6}{M_V^6} -\frac{2M_{V'}^4}{M_V^4} \right) \ln \left( \frac{M_{V'}^2}{M_A^2} -1\right) +\left[ -\frac{3M_A^{10}}{M_V^{10}} +\frac{14M_A^8}{M_V^8} -\frac{15M_A^6}{M_V^6} \right. \nonumber \\
&& \left. -\frac{2M_A^2}{M_V^2} +\left(2-\frac{M_A^2}{M_V^2} \right) \left(-\frac{M_A^8M_{V'}^2}{M_V^{10}}+\frac{9M_A^4M_{V'}^2}{M_V^6} -\frac{2M_A^2M_{V'}^2}{M_V^4} \right) \right] \ln \left( 1-\frac{M_V^2}{M_A^2} \right) \nonumber \\
&&+\left(1-\frac{M_{V'}^2}{M_V^2} \right) \left( \frac{2M_{V'}^6}{M_V^6} +\frac{2M_A^2}{M_V^2} +\frac{6M_A^2M_{V'}^2}{M_V^4} -\frac{5M_A^2M_{V'}^4}{M_V^6} \right) \ln \frac{M_V^2}{M_A^2} \, , \\
\nonumber \\
\xi_{P\pi}^{(1)} &=& \left(1-\frac{M_V^2}{M_{V'}^2}\right) \left( \frac{M_P^4}{M_V^4}+\frac{M_{V'}^2M_P^4}{M_V^6}-\frac{4M_{V'}^2M_P^2}{M_V^4}+\frac{M_{V'}^2}{M_V^2}+\frac{M_{V'}^4}{M_V^4} \right) \nonumber \\
&& + \left(1-\frac{M_P^2}{M_{V'}^2}\right)^2 \left( -\frac{M_P^2}{M_V^2}+\frac{3M_{V'}^2M_P^2}{M_V^4}-\frac{2M_{V'}^2}{M_V^2} \right) \ln \left( \frac{M_{V'}^2}{M_P^2}-1\right) \nonumber \\
&& +\left(1-\frac{M_P^2}{M_V^2} \right)^2 \left( -\frac{3M_P^2}{M_V^2} +\frac{2M_{V'}^2}{M_V^2} +\frac{M_{V'}^2M_P^2}{M_V^4} \right) \ln \left( 1-\frac{M_V^2}{M_P^2} \right) \, , \\
\nonumber \\
\xi_{P\pi}^{(2)} &=& \left(1-\frac{M_{V'}^2}{M_V^2}\right) \left( -1-\frac{2M_P^4}{M_V^4} +\frac{2M_P^2}{M_V^2}+\frac{2M_{V'}^2M_P^2}{M_V^4} -\frac{M_{V'}^4}{M_V^4} \right) \nonumber \\
&& +\left(1-\frac{M_{V'}^2}{M_V^2} \right) \ln \frac{M_V^2}{M_P^2} +\left(1-\frac{M_P^2}{M_V^2} \right)^2 \left(-1-\frac{2M_P^2}{M_V^2} +\frac{3M_{V'}^2}{M_V^2} \right) \ln \left(1-\frac{M_V^2}{M_P^2} \right) \nonumber \\
&& +\left( \frac{M_P^2}{M_V^2} -\frac{M_{V'}^2}{M_V^2} \right)^2 \left( -3+\frac{2M_P^2}{M_V^2} +\frac{M_{V'}^2}{M_V^2} \right) \ln \left( \frac{M_{V'}^2}{M_P^2} -1 \right) \, , \\
\nonumber \\
\xi_{V\pi}^{(1)} & = &
\Frac{6M_V^2}{M_A^2}\!+\!17\!-\!\Frac{6 M_A^2}{M_V^2} \!+\!\Frac{M_A^4}{M_V^4}
+\!\left(\Frac{6M_V^4}{M_A^4}\!+\!\Frac{14 M_V^2}{M_A^2} \!-\!15\! +\!\Frac{M_A^4}{M_V^4}\right)
   \ln{\left(\Frac{M_A^2}{M_V^2}-1\right)}
\, ,  \\
\nonumber \\
\xi_{V\pi}^{(2)} & = &
\Frac{M_A^6}{M_V^6}\!-\!\Frac{5M_A^4}{M_V^4}\!+\!\Frac{9 M_A^2}{M_V^2} +4
+ \! \left( \Frac{2 M_A^6}{M_V^6}\!-\!\Frac{7 M_A^4}{M_V^4}+7+\Frac{4M_V^2}{M_A^2}\right)
   \ln{\left(\Frac{M_A^2}{M_V^2}-1\right)}
\, ,  \\
\nonumber \\
\xi_{S\pi}^{(1)} &= &
3 \left(1-\Frac{M_S^2}{M_{A'}^2}\right)^2 \ln{\left(\Frac{M_{A'}^2}{M_S^2} -1\right)}
+\Frac{3M_S^2}{M_{A'}^2} -\Frac{9}{2} +\Frac{M_{A'}^2}{M_S^2}
\, ,  \\
\nonumber \\
\xi_{S\pi}^{(2)} & = &
2-\Frac{2 M_{A'}^2}{M_S^2}+\Frac{M_{A'}^4}{M_S^4}
 + \left( \Frac{2M_S^2}{M_{A'}^2} -3 +\Frac{M_{A'}^4}{M_S^4}\right)
 \ln{\left(\frac{M_{A'}^2}{M_S^2}-1\right)}
  - \Frac{M_{A'}^4}{M_S^4}\ln{\Frac{M_V^2}{M_S^2}}
\,.
\end{eqnarray}

Finally, we present the required functions for the determination of the chiral couplings $L_{10}^r(\mu)$ and $C_{87}^r(\mu)$, $\chi^{(1)}_{m_1m_2}$ and $\chi^{(2)}_{m_1m_2}$ respectively. They depend only on the resonance masses.
\begin{eqnarray}
\chi^{(1)}_{A\pi}&=& \left(1-\frac{M_V^2}{M_{V'}^2}\right) \left[-\frac{2M_{V'}^4}{M_V^4}-\frac{3M_A^2M_{V'}^2}{M_V^4}
 -\frac{2M_A^4}{M_V^4} \left(3+\frac{14M_{V'}^2}{M_V^2} \right) +\frac{M_A^6}{M_V^6}\nonumber \right. \\
&&\left. \times \left(-22+\frac{13M_V^2}{M_{V'}^2}+\frac{29M_{V'}^2}{M_V^2} \right) +2\left(1+\frac{M_V^2}{M_{V'}^2}\right)
 \frac{M_A^8}{M_V^8} \left( 5-\frac{3M_V^2}{M_{V'}^2}-\frac{3M_{V'}^2}{M_V^2}\right) \right] \nonumber \\
&& -\frac{2M_A^2}{M_V^2} \left( \frac{3M_A^8M_{V'}^2}{M_V^{10}} -\frac{5M_A^8}{M_V^8} +\frac{28M_A^6}{M_V^6} -\frac{16M_A^6M_{V'}^2}{M_V^8}+\frac{21M_A^4M_{V'}^2}{M_V^6}\right. \nonumber \\
&&\left.    -\frac{45M_A^4}{M_V^4} +\frac{14M_A^2}{M_V^2} +\frac{6M_A^2M_{V'}^2}{M_V^4} +2 -\frac{8M_{V'}^2}{M_V^2} \right) \ln \left( 1-\frac{M_V^2}{M_A^2} \right) \nonumber \\
&&  +\frac{2M_A^2}{M_V^2}\left(1-\frac{M_A^2}{M_{V'}^2}\right)^3  \left( -1-\frac{3M_A^2}{M_{V'}^2}+\frac{5M_A^2}{M_V^2}-\frac{5M_{V'}^2}{M_V^2} \right) \ln \left( \frac{M_{V'}^2}{M_A^2} -1\right) \, , \label{LEApi1} \\
&& \nonumber \\
\chi^{(2)}_{A\pi}&=&\left(1-\frac{M_V^2}{M_{V'}^2}\right)\left[ \frac{M_{V'}^2}{3M_V^2}
+\frac{2M_A^2}{3M_V^2} \left(16-\frac{31M_{V'}^2}{M_V^2} \right) +\frac{M_A^4}{3M_V^4} \left(112-\frac{65M_V^2}{M_{V'}^2} \right.\right.\nonumber \\
&& \left.  -\frac{155M_{V'}^2}{M_V^2} \right)
+\frac{2M_A^6}{M_V^6} \left( -11 +\frac{21M_{V'}^2}{M_V^2} -\frac{11M_V^2}{M_{V'}^2} +\frac{11M_V^4}{M_{V'}^4} \right) +\frac{4M_A^8}{M_V^8}\left( 1+\frac{M_V^4}{M_{V'}^4}\right. \nonumber \\
&& \left. \left. -\frac{2M_V^6}{M_{V'}^6}+\frac{M_{V}^2}{M_{V'}^2} -\frac{2M_{V'}^2}{M_V^2} \right) \right]
+\frac{2M_A^2}{M_V^2} \left( \frac{6M_A^8}{M_V^8} -\frac{4M_A^8M_{V'}^2}{M_V^{10}}+ \frac{23M_A^6M_{V'}^2}{M_V^8}\right. \nonumber \\
&& \left. -\frac{35M_A^6}{M_V^6} +\frac{60M_A^4}{M_V^4} -\frac{36M_A^4M_{V'}^2}{M_V^6} +\frac{M_A^2M_{V'}^2}{M_V^4} -\frac{21M_A^2}{M_V^2} +\frac{10M_{V'}^2}{M_V^2} -4 \right)  \nonumber \\
&&  \times \ln \left(1-\frac{M_V^2}{M_A^2} \right)  +2\left(1-\frac{M_A^2}{M_{V'}^2}\right)^3 \left( \frac{M_{V'}^2}{M_V^2} -\frac{7M_A^2}{M_V^2} -1 +\frac{6M_A^4}{M_V^2M_{V'}^2} +\frac{M_A^2}{M_{V'}^2} \right. \nonumber \\
&&  \left.  -\frac{4M_A^4}{M_{V'}^4} \right)  \ln \left( \frac{M_{V'}^2}{M_A^2} -1\right) \,, \label{LEApi2}\\
&& \nonumber \\
\chi^{(1)}_{P\pi}&=& 2\left(1-\frac{M_V^2}{M_{V'}^2}\right) \left( \frac{M_P^4}{M_V^2M_{V'}^2} -\frac{M_P^4}{M_V^4} +\frac{M_P^4 M_{V'}^2}{M_V^6} -\frac{M_P^2}{M_V^2} -\frac{M_P^2 M_{V'}^2}{M_V^4} +\frac{M_{V'}^2}{M_V^2} \right) \nonumber \\
&&   +\left(1-\frac{M_P^2}{M_V^2} \right)^2 \left(1-\frac{4M_P^2}{M_V^2} +\frac{2M_P^2M_{V'}^2}{M_V^4} +\frac{M_{V'}^2}{M_V^2} \right)\ln \left(1-\frac{M_V^2}{M_P^2} \right) \nonumber \\
&&   +\left(1-\frac{M_P^2}{M_{V'}^2} \right)^2 \left( -1-\frac{2M_P^2}{M_{V'}^2}+\frac{4M_P^2}{M_V^2} -\frac{M_{V'}^2}{M_V^2} \right) \ln \left( \frac{M_{V'}^2}{M_P^2} -1\right) \,, \label{LEPpi1} \\
&& \nonumber \\
\chi^{(2)}_{P\pi}&=&\left(1-\frac{M_V^2}{M_{V'}^2} \right) \left[ \left(1+\frac{M_V^2}{M_{V'}^2} \right) \frac{M_P^4}{M_V^4} \left( -5+\frac{3M_V^2}{M_{V'}^2} +\frac{3M_{V'}^2}{M_V^2} \right) +\frac{M_P^2}{2M_V^2} \Bigg( 10 \right.\nonumber \\
&& \left. \left.-\frac{9M_V^2}{M_{V'}^2}-\frac{9M_{V'}^2}{M_V^2} \right) +\frac{M_{V'}^2}{M_V^2} +1 \right] + \left(1-\frac{M_P^2}{M_V^2} \right)^2 \left( 2-\frac{5M_P^2}{M_V^2} +\frac{3M_P^2M_{V'}^2}{M_V^4} \right) \nonumber \\
&& \times    \ln \! \left( 1\!-\!\frac{M_V^2}{M_P^2} \right)\! +\!\left(1\!-\!\frac{M_P^2}{M_{V'}^2} \right)^2\! \left( -2\!+\!\frac{5M_P^2}{M_{V'}^2}\!-\!\frac{3M_P^2M_V^2}{M_{V'}^4} \right)\! \ln \!\left( \frac{M_{V'}^2}{M_P^2}\!-\!1\right)\!,  \label{LEPpi2}\\
&& \nonumber \\
\chi^{(1)}_{V\pi}&=& \!\left( 7 \!-\!\Frac{30 M_V^2}{M_A^2}\!+\!\Frac{21 M_V^4}{M_A^4} \!+\!\Frac{8 M_V^6}{M_A^6} \right)\! \ln{\!\left(\Frac{M_A^2}{M_V^2}\!-\!1\right)} 
 \!+\!\Frac{8 M_V^4}{M_A^4} \!+\!\Frac{25 M_V^2}{M_A^2} \! -\!\Frac{101}{6}\!+\!\Frac{M_A^2}{M_V^2} ,  \label{LEVpi1}\\
&&\nonumber \\
\chi^{(2)}_{V\pi} &=&\left( -\Frac{M_A^2}{M_V^2} +14 -\Frac{45 M_V^2}{M_A^2}+\Frac{28 M_V^4}{M_A^4}  +\Frac{10 M_V^6}{M_A^6}\right)\, \ln{\left(\Frac{M_A^2}{M_V^2}-1\right)} \nn  \\
&& +\Frac{10 M_V^4}{M_A^4} +\Frac{33 M_V^2}{M_A^2}  -\Frac{83}{3} +\Frac{10 M_A^2}{3 M_V^2} \, , \label{LEVpi2} \\
&& \nn \\
\chi^{(1)}_{S\pi}&=&  \left(-1 +\Frac{6 M_S^2}{M_{A'}^2} -\Frac{9 M_S^4}{M_{A'}^4}+\Frac{4 M_S^6}{M_{A'}^6} \right) \ln{\left(\Frac{M_{A'}^2}{M_S^2}-1\right)} 
  +\Frac{4 M_S^4}{M_{A'}^4}-\Frac{7 M_S^2}{M_{A'}^2} +\Frac{17}{6} \, , \label{LESpi1}\\
&& \nn \\
\chi^{(2)}_{S\pi}&=& \! \left(-2 \!+\!\Frac{9 M_S^2}{M_{A'}^2}\! -\!\Frac{12 M_S^4}{M_{A'}^4}\!+\!\Frac{5 M_S^6}{M_{A'}^6}  \right)\! \ln{\!\left(\Frac{M_{A'}^2}{M_S^2}\!-\!1\right)\!}  
  \!+\!\Frac{5 M_S^4}{M_{A'}^4}\!-\!\Frac{19 M_S^2}{2 M_{A'}^2} \!+\!\Frac{14}{3}   \!-\!\Frac{M_{A'}^2}{4M_S^2} \,. \label{LESpi2}
\end{eqnarray}

\section{Dispersive relation}\label{ap:C}

In the purely perturbative calculation (without Dyson resummations) and under the Single Resonance Approximation, the two-point function at next-to-leading order in the $1/N_C$ expansion reads as:
\begin{eqnarray}
\Pi(t)&=& \Frac{D(t)}{\left(M_R^2  -  t\right)^2} \, ,
\end{eqnarray}
where $M_R$ is the mass of the corresponding resonance in the $s$--channel, and $D(t)$ is an analytical  function except for the unitarity logarithmic branch (without poles).

\begin{figure}
\begin{center}
\includegraphics[angle=0,clip,width=5cm]{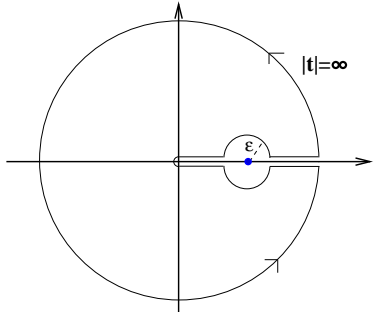}
\caption{\label{fig.circuito}
Integration circuit.}
\end{center}
\end{figure}

In order to recover the correlator, the complex integration in the circuit of Figure~\ref{fig.circuito} is performed:
\begin{eqnarray}
\Pi(s) & = & \Frac{1}{2\pi i} \, \oint \, \  \mathrm{d}t\; \Frac{\Pi(t)}{t -  s }  \, .
\end{eqnarray}
If it is assumed that $|\Pi(t)|\to 0$ when $|t|\to \infty$, the contribution from the external circle of the circuit is zero and it is found that:
\begin{eqnarray}\label{eq.master}
\Pi(t)&=& \sum_{m_1,m_2} \Pi(t)|_{m_1,m_2}
-  \Frac{\mbox{Re}D'(M_R^2)}{M_R^2-q^2}  +  \Frac{\mbox{Re}D(M_R^2)}{\left(M_R^2-q^2\right)^2} \, ,
\end{eqnarray}
with $D'(t)\equiv \Frac{d}{dt}D(t)$ and with the different contributions
of each two-meson absorptive cut given by the dispersive integral,
\begin{eqnarray}
\label{DeltaPi}
\Pi(s)|_{m_1,m_2} &= & \lim_{\epsilon\to 0} \left[ \int_0^{M_R^2-\epsilon}
\!\!\! \mathrm{d}t\; \Frac{1}{\pi}\Frac{\mbox{Im}\Pi(t)|_{m_1,m_2}}{t\, -\, s}
\, + \,
\int_{M_R^2+\epsilon}^\infty \!\!\! \mathrm{d}t\;
\Frac{1}{\pi}\Frac{\mbox{Im}\Pi(t)|_{m_1,m_2}}{t\, -\, s}
\right. \nonumber
\\
&&\qquad \left. \,-\, \Frac{2}{\pi\epsilon}
\, \lim_{t\to M_R^2}
\left\{(M_R^2-t)^2\, \Frac{\mbox{Im}\Pi(t)|_{m_1,m_2}}{t\,-\,s}
\right\}\,\,
\right]\, ,
\end{eqnarray}
where $M_R$ is the mass of the intermediate resonance produced in the
$m_1,m_2$ form-factor.



\hspace*{0cm} From  Eq.~(\ref{eq.master}),  one notices that, as soon as the value of
the real part of $D(t)$ and its first derivative are fixed at $M_R^2$, the whole correlator becomes fixed
by them and the spectral function at $t\neq M_R^2$.
This corresponds to providing a renormalization prescription for the corresponding coupling and resonance mass.

The fact that the spectral function vanishes at infinite momentum ensures that there are no terms
of the form $\Pi(t)\sim t^{m}\, \ln{(-t)}$, with $m\geq 0$. Furthermore, the polynomial terms
$\Pi(t)\sim t^{m}$ with $m\geq 0$ must be also identically zero in order to keep
$\Pi(t)\rightarrow0$ at $|t|\rightarrow \infty$.  Hence, the expression in Eq.~(\ref{eq.master}),
is the general expression for the correlator within the SRA. The inclusion of higher resonances can be performed in a straightforward way.

This means that although the presence of $\cO(p^4)$ $\chi$PT operators with NLO couplings in $1/N_C$,
$\widetilde{L}_i$,  is not forbidden by the symmetry, the QCD short-distance behaviour imposes that,
in our realization,  they do not get renormalized, as suggested in Ref.~\cite{CP:02},
and they do not contribute to the observable at the end of the day.
This provides a further understanding to the lack of running found in the $\widetilde{L}_i$ couplings
in the one-loop analysis of the generating functional
performed in Ref.~\cite{RPP:05} after imposing the high-energy constraints.

\subsection{Diagrammatic Calculation}
For sake of simplicity we will refer now just to the vector correlator although
the extension to other channels is straightforward. At tree-level, it is found that
\begin{eqnarray}
\Pi_{_{VV}}(t)&=& \Frac{2 F_V^2}{M_V^2 - t} \, .
\end{eqnarray}
The resonance parameters $F_V$ and $M_V$ get renormalized  at the next-to-leading order in $1/N_C$
($F_V=F_V^{\, r}+\delta F_V$ and $M_V^2=M_V^{r\,\,2}+\delta M_V^2$)
in order to cancel the ultraviolet divergences from the one-loop diagrams:
\begin{eqnarray}
\Pi_{_{VV}}(t)|_{\mathrm{tree}}\, =&\mbox{}\hspace{-3.5cm} \Frac{2\,F_V^{r\,2}}{M_V^{r\,\,2} - t}
+ \Frac{4\,F_V^{r}\,\delta F_V}{M_V^{r\,\,2} - t}
- \Frac{2\,F_V^{r\,2}\,\delta M_V^2}{\left(M_V^{2\,\,r} - t\right)^2}  + \cO\!\left(\Frac{1}{N_C}\right) , 
\label{tree} \\
\Pi_{_{VV}}(t)|_\mathrm{1-loop}&= \Frac{D(t)|_\mathrm{1-loop}}{\left(M_V^{r\,\,2} - t\right)^2}
\; =
\displaystyle{\sum_{m_1,m_2}}\Pi(t)|_{m_1,m_2}
+ \Frac{c_1 +  \gamma_1\, \lambda_\infty}{M_V^{r\,\,2} - t}
+ \Frac{c_2 +  \gamma_2 \, \lambda_\infty}{\left(M_V^{r\,\,2} -t \right)^2}\, ,\qquad\mbox{} \label{loop}
\end{eqnarray}
where  $c_{1,2}$ and $\gamma_{1,2}$ are constants determined by the one-loop calculation. Taking into account Eq.~(\ref{eq.master}), one gets
\begin{eqnarray}
c_1+\gamma_1\, \lambda_\infty &=& -\mbox{Re}\left\{D\,' (M_V^{r\,\,2})|_\mathrm{1-loop} \right\} \,,\nonumber \\
 c_2+\gamma_2\,\lambda_\infty& =& \mbox{Re}\left\{ D(M_V^{r\,\,2})|_\mathrm{1-loop}\right\} \,.
\end{eqnarray}

All the relevant  ultraviolet divergences are shown in Eq.~(\ref{loop}).
As mentioned before, the polynomial divergences
$\Pi_{_{VV}}(t)\sim \gamma_{-m}\, t^m\,  \lambda_\infty $
cannot produce any contribution at the end of the day, so they exactly cancel at  any energy.
Once again,   considering well behaved correlators --and therefore form factors-- at large energies
is crucial.

The renormalization procedure through the $F_V$ and $M_V$ counter-terms gives
\begin{eqnarray}
4 F_V^{r}\delta F_V + \gamma_1\,\lambda_\infty  &=&  0 \, , \nonumber\\
 -2\,F_V^{r\,\, 2} \, \delta M_V^{r\,\,2} + \gamma_2\,\lambda_\infty  &=& 0 \, .
\end{eqnarray}

The renormalized amplitude up to next-to-leading order in the $1/N_C$ expansion shows the general structure
\begin{eqnarray}
\Pi_{_{VV}}(t)& =&
\sum_{m_1,m_2}\Pi(t)|_{m_1,m_2}
+\Frac{2 F_V^{r\,\,2} + c_1}{M_V^{r\,\,2}  - t} +\Frac{c_2}{\left(M_V^{r\,\,2} - t \right)^2}\, .
\end{eqnarray}
The unknown subtraction constants $c_1$ and $c_2$ can be absorbed in a redefinition of
$F_V^{r}$ and $M_V^{r}$. One can set them to zero, {\it i.e.}, $c_1=c_2=0$,
and $F_V^r$ and $M_V^r$ result then renormalization-scale independent.

\subsection{Contribution from High-Mass Absorptive Cuts}

Because of the approximation of neglecting intermediate states with two resonances, made in section~4.4, it is convenient to analyse the  effect on the $\chi$PT couplings of absorptive cuts with higher and higher production thresholds. When the threshold $\Lambda_{th}^2$ is above the resonance mass $M_R^2$, one finds for the low energy limit $q^2\ll \Lambda_{th}^2$,
\begin{equation}
\Pi(q^2)|_\rho \,=\, \Int_{\Lambda_{th}^2}^\infty \mathrm{d}t\; \Frac{1}{\pi} \Frac{\mbox{Im}\Pi(t)}{t-q^2}
\,=\, \displaystyle{\sum_{n=0}^\infty}
\left(\Frac{q^2}{\Lambda_{th}^2}\right)^n \Int_{1}^\infty \mathrm{d}x \;
\Frac{1}{\pi} \Frac{\mbox{Im}\,\Pi(x\cdot \Lambda_{th}^2)}{x^{n+1}}  \, .
\end{equation}
The contributions become smaller and smaller as the value of the production threshold $\Lambda_{th}^2$ is increased, supporting  the approximation in section~4.4.

On the other hand, in the deep euclidean region $Q^2=-q^2\gg \Lambda_{th}^2$, one gets
\begin{equation}
\left|\Pi(q^2)|_\rho\right| \,\leq \, \Frac{1}{Q^2}\, \Int_{\Lambda_{th}^2}^\infty \mathrm{d}t\; \Frac{1}{\pi} \left|\mbox{Im}\Pi(t)\right| \, ,
\end{equation}
which becomes smaller and smaller as $\Lambda_{th}^2$ is increased.

Note the importance of a well-behaved spectral function, that is, $\mathrm{Im} \Pi(t)\rightarrow \infty$ at $|t|\rightarrow \infty$, in order to be able to use the expressions of this appendix.


\end{document}